\RequirePackage{snapshot}
\documentclass[final]{ulthesis}
\usepackage{url}
\usepackage{multicol}

\usepackage[pdfauthor={Ahmed Elsayed},
            pdftitle={A Biometric Sensor Network to Enable Real-Time Measurement and Monitoring of Individual Student Engagement in STEM Lecture Environments},
            pdfsubject={Electrical Engineering},
            pdfkeywords={student engagement, biometric sensor network, computer Vision, CV, machine learning},
            pdfproducer={Latex with hyperref},
            pdfcreator={latex->dvips->ps2pdf},
            pdfpagemode=UseOutlines,
            bookmarksopen=true,
            bookmarksnumbered=true]{hyperref}

\hypersetup{
    colorlinks=true,
    linkcolor=black,
    filecolor=black,      
    urlcolor=black
}
\usepackage{memhfixc}

\usepackage[pdftex]{graphicx}
\usepackage{verbatim}   
\usepackage{color}      
\usepackage{hyperref}  
\usepackage{pgffor}
\usepackage{pdfpages}
\usepackage{wrapfig}
\usepackage{longtable}
\usepackage{multirow}
\usepackage{multicol}
\usepackage{array}
\usepackage{booktabs}   
\usepackage{tabularx}   
\usepackage{rotating}
\usepackage{pdflscape}
\usepackage{graphicx}
\usepackage{tikz}
\usetikzlibrary{shapes.geometric, arrows, positioning}

\usepackage{amsmath}    
\usepackage{listings}   

\usepackage[labelsep=period, labelfont=bf]{caption} 

\usepackage{chngcntr}   
\usepackage{relsize}    
\usepackage{subcaption} 

\makeatletter
\renewcommand\@memmain@floats{%

\counterwithout{figure}{chapter}
\counterwithout{table}{chapter}
\counterwithout{equation}{chapter}}

\def\babar{\mbox{\slshape B\kern-0.3em{\smaller[2] A}\kern-0.1em B\kern-0.3em{\smaller[2] A\kern-0.2em R}} }

\begin{document}

\def \apjl{Astrophysical Journal Letters }
\def \apjs{Astrophysical Journal Supplement Series }
\def \apj{Astrophysical Journal }
\def \aaps{Astronomy and Astrophysics Supplement }
\def \apss{Astrophysics and Space Science }
\def \mnras{Monthly Notices of the Royal Astronomical Society }
\def \aap{Astronomy and Astrophysics }
\def \aj{Astronomical Journal }
\def \pasp{Publications of the Astronomical Society of the Pacific }
\def \pasj{Publications of the Astronomical Society of Japan }
\def \ao{Applied Optics }


\author{Ahmed Elsayed}

\authordegree{B.S. in Electrical and Computer Engineering, 2016}

\title{A Biometric Sensor Network to Enable Real-Time Measurement of Individual Student Engagement in STEM Lecture Environments}

\dissertationdegree{Master of Science}

\dissertationdiscipline{Electrical Engineering}

\dissertationdate{December 2025}

\approvaldate{December 11, 2025}

\school{J.B. Speed School of Engineering}

\department{Department of Electrical and Computer Engineering}

\advisor{Michael McIntyre, Ph.D., Advisor}
\firstmember{Aly Farag, Ph.D.}
\secondmember{Thomas Tretter, Ph.D.}
\thirdmember{John Naber, Ph.D.}


\frontmatter

\maketitle


    


\begin{acknowledgments}

I would like to express my deepest gratitude to my mother for her unwavering love, sacrifices, and prayers—her support has been the foundation of everything I have accomplished. I am also profoundly grateful to my brother, sisters, cousins, nephews, and my entire family in Egypt, who have continually encouraged me and made countless sacrifices so I could pursue my academic goals. I extend my sincere appreciation to my friends, my colleagues at the CVIP Lab, and the Egyptian community in Louisville for their companionship, support, and encouragement throughout this journey.

I am especially grateful to Dr. Aly Farag for allowing me to join his research group at the CVIP Lab, and to Ahmed Shalaby for his continuous support and for recommending me for an opportunity at Apple. I am also grateful to Ahmed Abdelkawy for the many fruitful discussions and collaborative efforts during our time working together in the CVIP Lab, and to Dr. Michel MacIntyre for his helpful support.

\end{acknowledgments}



\begin{dissertationabstract}

Student engagement (SE) is a critical predictor of academic performance and retention in STEM education, yet existing measurement approaches are often intrusive, manually intensive, or unsuitable for real-time classroom use. This thesis proposes a novel \textit{Biometric Sensor Network} (BSN) designed to enable real-time measurement and continuous tracking of individual student engagement in STEM classroom environments. The system enables capturing of behavioral, emotional, and cognitive indicators through camera-based sensing while preserving ethical and privacy constraints.

To measure these indicators unobtrusively and ethically, we propose a BSN composed of \textit{Student Processing Units} (SPUs) that function as distributed sensing nodes. The network is explicitly designed to satisfy five objectives: it must be \textbf{non-intrusive}, \textbf{non-invasive}, \textbf{non-stigmatizing}, \textbf{real-time}, and \textbf{automatic}, while ensuring rigorous protection of student data security and privacy.

Each SPU supports two operational modes: (i) a \textit{dataset-collection mode}, in which raw student video is temporarily recorded to construct a private SE dataset for model training and validation, and (ii) an \textit{analysis mode}, in which the SPU performs real-time inference on 10-second video segments without storing or transmitting raw frames. In this analysis role, each SPU enables fully on-device processing---including face detection, gaze estimation, and affective analysis---ensuring that no identifiable video data leaves the device. A secure backend infrastructure manages device authentication, session orchestration, and encrypted data ingestion. The full system integrates hardware design, computer-vision pipelines, wireless networking, security protocols, and session-level data management.

A series of evaluations demonstrates that the proposed SPU meets the required performance benchmarks, including sustaining multi-device data bandwidth, maintaining sub-second clock synchronization accuracy, executing moderate CV/ML inference workloads in real time, and achieving more than two hours of battery-powered operation. The system significantly improves upon a previous baseline developed in the CVIP Lab in terms of security, autonomy, compute capability, and data reliability.

Overall, this work provides a complete, deployable BSN architecture that enables real-time measurement and tracking of individual student engagement during STEM classroom lectures while adhering to the strict ethical, privacy, and operational requirements of real-world educational environments.

\end{dissertationabstract}


\tableofcontents* \clearpage
\listoftables \clearpage
\listoffigures \clearpage



\mainmatter

\chapter{Introduction}
\section{Introduction}
Science, technology, engineering, and mathematics (STEM) majors play a crucial role in the economic development of industrialized nations \cite{Carnevale2015, Rothwell2013}. However, many of these countries continue to face challenges in producing sufficient numbers of STEM-trained graduates \cite{Beach2024, Malcom2016, Kelly2013}. For example, in the United States, projections suggest that the number of STEM degree earners would need to increase by roughly 33\% to meet future workforce demands \cite{Olson2012}. The major cause of this shortfall is the high attrition rate among students pursuing STEM majors. Studies show that in the United States, 50–60\% of students who enter college intending to complete a STEM degree ultimately either switch to a non-STEM field or leave college without earning a degree, despite being academically capable \cite{Chen2015, Chen2013, Olson2012}. Poor performance in foundational STEM “gateway” courses is a leading factor in student attrition from STEM majors \cite{Aulck2017, Chen2015}. Additional contributing factors include a limited sense of belonging, insufficient exposure to authentic research experiences, and the persistent underrepresentation of minority groups in STEM \cite{highschool2019, Rozek2019, Malcom2016, Theobald2020}.

High attrition rates in STEM majors call for urgent countermeasures, and many strategies have been proposed in the literature. Promising approaches include active learning with peer support \cite{Feng2025, Clements2025}, corequisite math pathways \cite{Ran2025}, belonging interventions \cite{Walton2023}, STEM intervention programs \cite{Shortlidge2024}, and course-based undergraduate research experiences (CUREs) \cite{Bekkering2025, Broussard2025}. Although these initiatives differ in design, they share a common mechanism: fostering student engagement (SE)—behavioral, emotional, or cognitive—as a means of reducing attrition \cite{Wang2014, Fredricks2004}.

During the past four decades, SE has been the focus of extensive research, supported by large-scale data sources such as annual engagement surveys conducted across hundreds of higher education institutions \cite{EngReprot2020}. Many studies have shown that higher levels of SE are associated with improved academic performance, higher likelihood of degree completion, lower dropout rates, and reduced negative behaviors within academic settings \cite{Carini2006, FREDRICKS2015, Adnan2021}. These findings underscore the importance of SE in reducing attrition, making it crucial to gain insight into how students engage within courses to improve teaching and learning outcomes and to identify timely interventions for at-risk students. Indeed, engagement has been described as ``the holy grail of learning`` \cite{Sinatra2015}.

This work addresses the issue of high attrition rates among college students from STEM majors by proposing a novel and effective biometric sensor network that enables real-time measurement and tracking of individual SE during STEM classroom lectures. This BSN is effective in the sense that it can operate in a non-intrusive, non-invasive, non-stigmatizing,  automatic, and real-time way while adhering to the regulations and policies imposed by the University of Louisville (UofL) Institutional Review Board (IRB). Furthermore, this BSN leverages recent advances in imaging sensors and edge computing to enable the measurement of SE through state-of-the-art algorithms in computer vision (CV) and machine learning (ML).

The organization of this work is as follows. Chapter~2 provides the theoretical background of SE from an educational psychology perspective; it defines SE, describes how it is measured, and reviews recent methods in the literature relevant to this research. Chapter~3 presents the analysis and design of the proposed BSN, including system analysis and key technical design decisions. Chapter~4 describes the hardware and software implementations of the design introduced in Chapter~3. Chapter~5 presents the experiments and evaluation, demonstrating the effectiveness of the proposed BSN. Finally, Chapter~6 provides the conclusion and outlines potential future directions.

\chapter{Background}

The purpose of this chapter is to provide the necessary background for readers who do not have a foundation in educational psychology. We begin with the adopted definition of SE, followed by a discussion of the distinctions between engagement, flow, and attention, as the latter two constructs---although different from engagement---are closely related to it and are sometimes used interchangeably in the literature. Next, we review the various methods used to measure SE, highlighting the advantages and limitations of each approach.

\section{Student Engagement Definition}
\label{sec:engagement-definitions}
In educational psychology, SE is often one of the most misused and overgeneralized constructs~\cite{Azevedo2015}. Therefore, any research that aims to measure SE must begin with a clear and precise definition of what SE is, ensuring that the construct under investigation is conceptually well-grounded before discussing how it can be measured.

SE has been widely conceptualized as a meta-construct comprising three interrelated dimensions---behavioral, emotional, and cognitive engagement---as described in the seminal 
framework proposed by Fredricks \emph{et al}.~\cite{Fredricks2004}. There is no consensus among researchers on a precise definition, and recent studies have proposed additional dimensions such as social engagement~\cite{Wang2019} and agentic engagement~\cite{Reeve2011}.
\begin{wrapfigure}{r}{0.5\textwidth}
    \centering
    \vspace{-10pt}
    \includegraphics[width=0.48\textwidth]{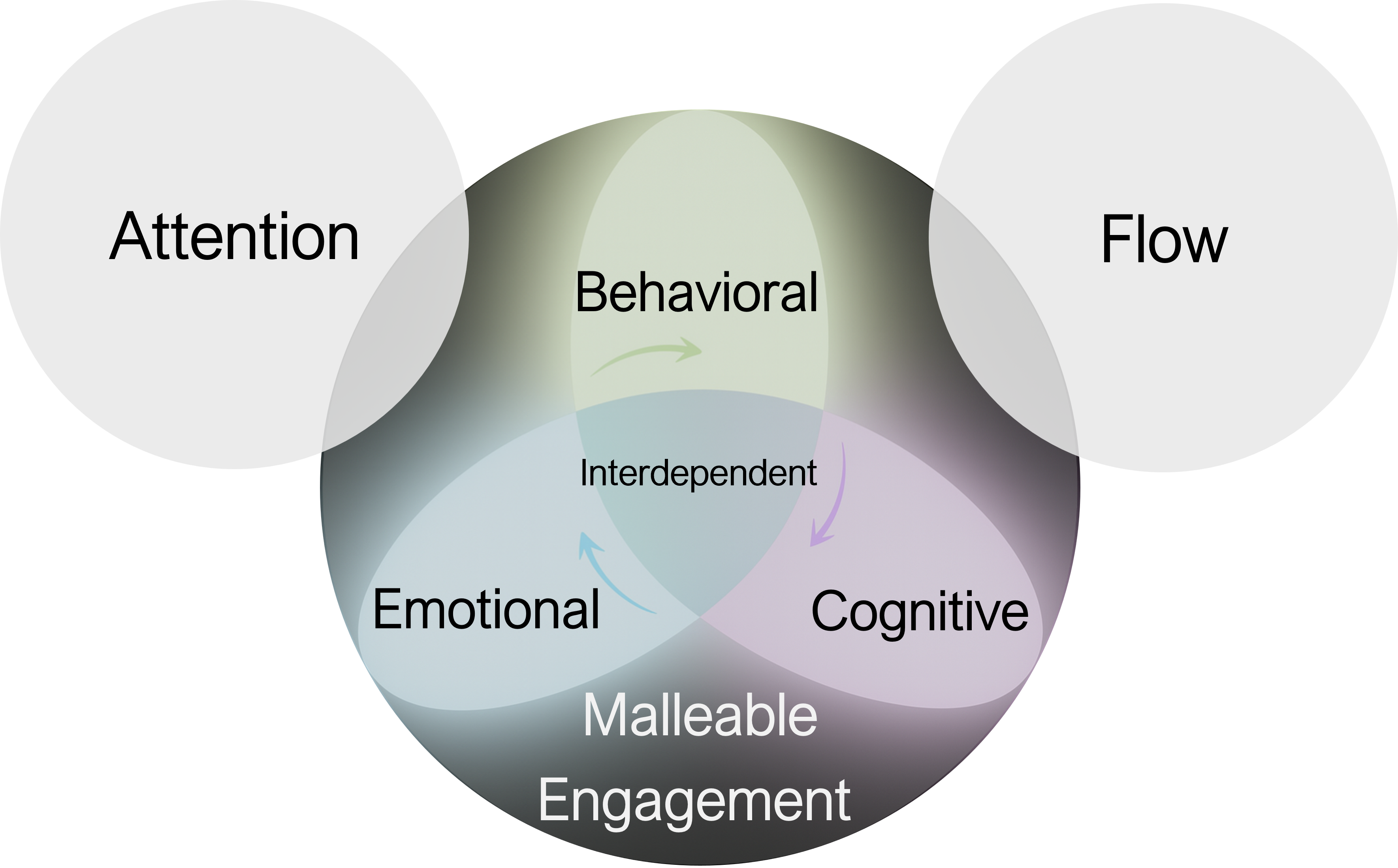}
    \caption{Venn diagram to represent engagement as a malleable construct in which each component is interdependent and cannot be completely disentangled from the other two. The diagram also shows that SE overlaps with related constructs such as attention and flow.}
    \label{fig:ENG_VENN_DIAG}
\end{wrapfigure}
However, it remains unclear whether these represent truly distinct dimensions or whether they can be subsumed under one of the three canonical components. In this work, we adopt the widely cited framework of Fredricks \emph{et al}.~\cite{Fredricks2004}, which conceptualizes engagement as a multidimensional construct consisting of three interrelated yet distinct dimensions: behavioral, emotional, and cognitive engagement. Within this framework, engagement is viewed as a malleable construct in which each component is interdependent and cannot be completely disentangled from the other two. This construct can be best described by the Venn diagram shown in Figure~\ref{fig:ENG_VENN_DIAG}.

\paragraph{Behavioral Engagement (BE).}
BE has been defined in multiple ways; in this study, we adopt the definition that frames it 
as active involvement in learning and academic tasks. This includes behaviors such as effort, persistence, concentration, attention, asking questions, and participating in class discussions~\cite{Fredricks2004}.

\paragraph{Emotional Engagement (EE).}
EE refers to students’ affective responses in the classroom, which encompass emotional reactions such as interest, boredom, happiness, sadness, anxiety, and anger~\cite{Skinner1993}.

\paragraph{Cognitive Engagement (CE).}
CE, similar to BE, has been defined in multiple ways, with conceptualizations emerging from research on both school engagement and learning and instruction~\cite{Fredricks2004}. In  this study, we adopt the definition proposed by Lamborn \emph{et al}.~\cite{Lamborn1992},  which emphasizes students’ psychological investment and effort in learning, describing CE as “the student’s psychological investment in and effort directed toward learning, understanding, and mastering the knowledge, skills, or crafts that the academic work is intended to promote.”

\section{Attention, Flow, and Engagement}

Attention and flow are two psychological constructs that, while distinct from SE, are closely related to it and help illuminate the mechanisms underlying students’ involvement in learning. Understanding these constructs is important because they provide insight into how learners allocate mental resources, regulate their focus, and experience immersion during academic activities.

Csikszentmihalyi~\cite{Csikszentmihalyi2014} describes attention as a limited form of psychic energy that regulates the stream of consciousness, with its allocation shaping one’s overall life experience. At the neurological level, attention comprises four sub-processes: working memory, top-down sensitivity control, competitive selection, and automatic filtering of salient stimuli~\cite{Knudsen2007}. Through selective competition, neural representations derived from salience filters and sensitivity control are chosen to enter working memory, where analysis, decision-making, and planning take place. This framework distinguishes between two types of attention: voluntary (top-down) and involuntary (bottom-up). Voluntary attention relies on sensitivity control, competitive selection, and working memory functioning within a stable feedback loop, whereas involuntary attention is primarily triggered by infrequent, salient stimuli occurring suddenly in time or space (e.g., the blare of an ambulance siren or the flashing lights of a police vehicle). Csikszentmihalyi~\cite{Csikszentmihalyi2014} argues that voluntarily 
focusing attention on a limited stimulus field is necessary to attain experiences that are subjectively and socially valued, whereas difficulty regulating voluntary attention can contribute to psychopathology.

Flow is described as a subjective, intrinsically enjoyable experiential state in which an individual becomes fully absorbed in a task, losing awareness of distractions and even the passage of time~\cite{Csikszentmihalyi1990}. In educational settings, the highest level of SE can be understood as the experience of flow sustained throughout a learning activity. For flow to occur, concentration, interest, and enjoyment must coexist within the activity (e.g., attending a lecture), and thus a measure of engagement can be constructed from these three variables~\cite{Shernoff2014}.

Although attention and flow are distinct constructs, they are closely connected to SE because they reflect core psychological processes that govern how students participate in learning. Attention determines how cognitive resources are allocated, enabling learners to focus on instructional content, persist in academic tasks, and resist distraction—functions that directly support behavioral and cognitive engagement. Flow, in contrast, represents an optimal experiential state in which attention is fully directed toward the task and is accompanied by heightened interest, enjoyment, and intrinsic motivation. Sustained voluntary attention is a prerequisite for entering the flow state, and the experience of flow itself reflects a high-intensity form of engagement. Together, attention and flow help explain how learners invest effort, regulate emotion, and maintain involvement in academic activities, thereby reinforcing their conceptual connection to the canonical dimensions of SE.

\section{Engagement Measures}
\label{sec:eng_meas}
SE assessment in the literature can be classified into six main approaches: student self-report surveys, teacher ratings, observational measures, administrative data, experience sampling methods (ESM), and real-time measurements \cite{Fredricks2022}. The advantages and limitations of each approach are briefly outlined below, with a more detailed discussion devoted to real-time measurement, as our method belongs to this category.

\textit{Student self-report surveys} are the most widely used method. They are low-cost, scalable, and capture students’ subjective perceptions and links between contextual factors and engagement. However, they are limited to emotional and cognitive dimensions, provide only school- or classroom-level insights, treat engagement as a static
\begin{wrapfigure}{r}{0.5\textwidth}
    \centering
    \vspace{-10pt}
    \includegraphics[width=0.48\textwidth]{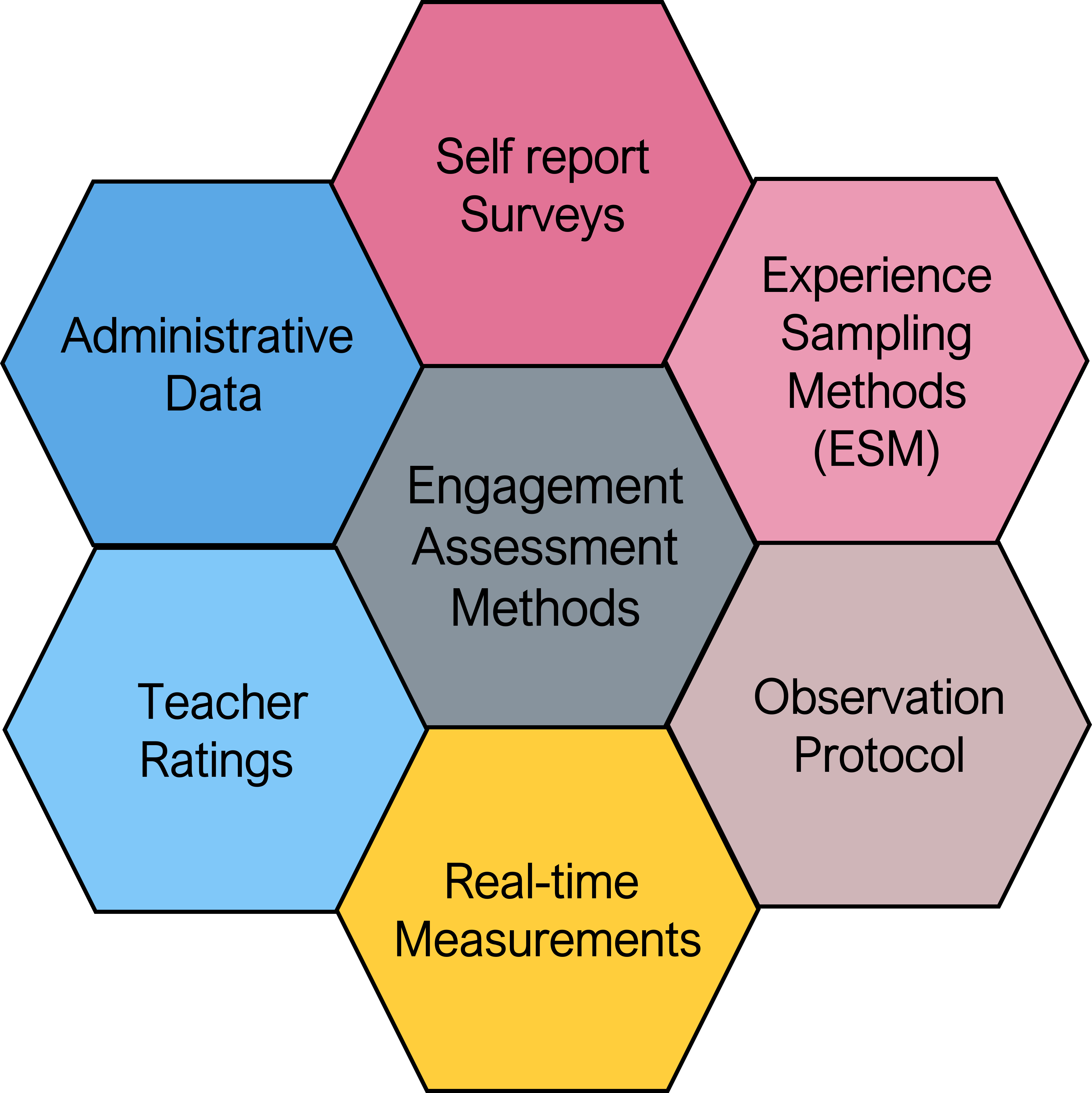}
    \caption{The six primary methods for measuring SE reported in the literature.}
    \label{fig:ENG_MEAS_METHODS}
\end{wrapfigure}
post-task construct, may not reflect actual behaviors or strategies, and are prone to social desirability bias \cite{Fredricks2022}.
\textit{Teacher ratings} rely on teachers to score students’ engagement across multiple indicators. They are simple, inexpensive, and scalable, but mainly capture behavioral engagement, while cognitive and emotional dimensions are harder to infer. They are also subject to bias from both student and teacher characteristics \cite{Fredricks2022}. 

\textit{Administrative data}, such as attendance and grades, measure behavioral engagement. Collected routinely on all students, they allow longitudinal tracking and early intervention. Yet, they conflate outcomes with indicators of engagement, are biased by student characteristics, and lack standardization across institutions \cite{Fredricks2022}.

\textit{ESP} is rooted in flow research \cite{Shernoff2022}. Students are randomly interpreted via ESM signals to fill out short surveys about their location, activities, behavior, and cognitive and affective responses. It offers time and context-dependent measures of student subjective experience, allowing data on engagement to be collected as it happens. These methods can be applied repeatedly to a large population, allowing for comparison of SE levels across time and context. The cons of this technique include the necessity of a high level of commitment from participants. suffers from participant fatigue, hasty completion exaggeration, and deliberation falsification \cite{Fredricks2022}.

\textit{Observational measures} involve trained coders who systematically watch students and apply a predefined coding scheme to record specific behavioral indicators within a fixed period of time. These measures can capture behavioral engagement at the student, group, or classroom levels, provide descriptions of both engagement and its surrounding context, and are generally less disruptive to classroom activities. However, they also require decisions regarding the sampling period and units of analysis; they are labor-intensive, limited to relatively small populations, susceptible to observer bias, and difficult to generalize across different educational contexts \cite{Fredricks2022}.

\begin{wrapfigure}{r}{0.5\textwidth}
    \centering
    \vspace{-10pt}
    \includegraphics[width=0.48\textwidth]{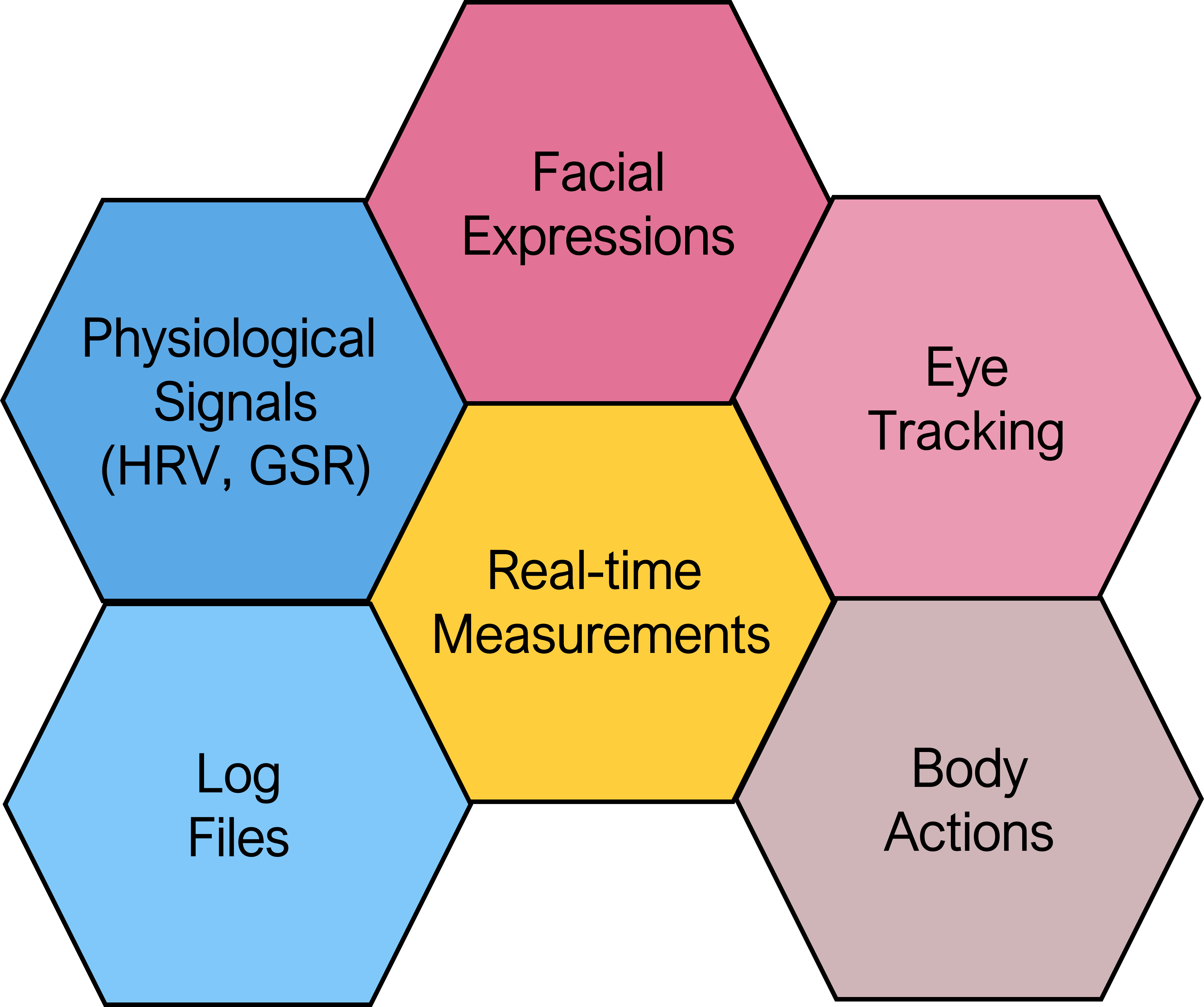}
    \caption{The five primary types of signals used as real-time indicators to assess SE as reported in the literature.}
    \label{fig:REAL_TIME_IND}
\end{wrapfigure}

\section{Real-Time Engagement Measures}
\label{sec:realtime_meas}
All methods discussed so far are limited by coarse temporal granularity, capturing engagement at low-resolution timescales or lacking automated data collection. These approaches are therefore unable to detect rapid fluctuations in SE or provide continuous, fine-grained measurement during learning activities.

In contrast, a more recent and promising approach, \textit{real-time measures}, has emerged, enabled by advances in CV, ML, sensor technologies, computer networks, and edge computing. This approach tracks SE dynamics at temporal scales as fine as seconds or minutes by relying on discrete and objective indicators.

In the literature, five main types of indicators are commonly used to assess SE: log files \cite{Gobert2015, Li2023}, eye tracking \cite{Carter2020}, facial expressions \cite{Tang2025, Alkabbany2023}, body actions \cite{Abdelkawy2024, Ashwin2019}, and physiological signals such as electroencephalogram (EEG), heart rate variability (HRV), and galvanic skin response (GSR) \cite{Kim2018}. Using real-time measures offers several advantages: it is objective and less prone to student or teacher biases, provides detailed information about how engagement changes over time, can be automated to reduce interruptions, and allows the collection of large amounts of data in a short period \cite{Fredricks2022}.

However, this approach is not without limitations. Real-time data collection requires technical expertise, is often conducted in small, structured tasks rather than dynamic classroom environments, raises privacy concerns, and can be costly in terms of data processing and storage. Furthermore, there is no consensus on the optimal temporal 
granularity for measuring engagement, and the resulting data must be carefully analyzed—typically using CV and ML algorithms—which may lack interpretability and require careful validation to produce meaningful results \cite{Fredricks2022}.

Log files are outside the scope of this work as they are typically used in e-learning environments. Physiological signals can be applied in classrooms, but often require sophisticated, intrusive, or invasive sensors and are affected by body movements or sweat, making them less suitable for our purposes. Eye tracking records learners’ gaze patterns to reveal visual attention and engagement with specific areas of interest (AOIs), such as the blackboard or teacher \cite{Carter2020}. It is closely related to cognitive load (CL) and is considered an important indicator of cognitive engagement \cite{Walter2021, Liu2025}.

Facial expressions (FE) can be described in terms of facial action units (FACS) \cite{Ekman1978} and can convey internal human emotions \cite{Ekman1993}. Many algorithms have been proposed for automatic recognition of facial expressions (FER) from camera images or videos, and comprehensive reviews of this research are available in \cite{Li2022, Corneanu2016}. FER has been explored as a promising tool to automatically measure students’ emotional engagement \cite{Fredricks2022}.

Body action refers to observable movements or postures of the human body, which can be analyzed to infer activities, intentions, or emotional states \cite{Huang2024, Dael2012}. In CV, body action recognition is defined as the process of labeling sequences of body movements captured through cameras or sensors into predefined action categories \cite{Aggarwal2011, Poppe2010}. This approach has been successfully applied to measure student behavioral engagement in classrooms by analyzing students’ actions during lectures \cite{Abdelkawy2024, Alkabbany2019}.

\section{Literature Review (zReal-time Methods)}
Many methods have been proposed in the literature to measure one or more dimensions of SE based on real-time indicators. In this section, only methods that rely on visual cues are discussed. Sheng et al. \cite{Sheng2025} introduced an improved YOLOv8-based detector to localize and classify student behaviors from single classroom images; however, their work stopped at behavior detection and did not link the detected behaviors to validated measures of engagement, nor did it evaluate whether behavior counts or temporal patterns could predict engagement levels. Nezami et al. \cite{MohamadNezami2020} constructed a private dataset of SE comprising frontal facial images of students interacting with a virtual world environment for science education. Engagement and disengagement labels were assigned by annotating images with both behavioral and emotional cues. Their deep learning model achieved a classification accuracy of 72.38\%; however, the approach treats engagement as a static construct that can be inferred from a single image. This assumption overlooks the inherently dynamic and temporal nature of engagement. Furthermore, the method is tailored to e-learning or virtual learning contexts and does not readily generalize to traditional classroom settings, where social interactions, group dynamics, and environmental complexity play a critical role in shaping engagement.

Singh et al. \cite{Singh2023} addressed key limitations in prior engagement research by introducing EngageNet, a large-scale dataset comprising 31 hours of video from 127 participants captured in diverse real-world conditions. Unlike earlier datasets that primarily conceptualized engagement as a static construct inferred from short clips, EngageNet incorporates behavioral, cognitive, and self-reported measures, providing a more nuanced and multidimensional view of engagement. The authors also established strong baseline models using facial action units, gaze, head pose, and transformer-based architectures, and validated their approach across both EngageNet and the EngageWild dataset to demonstrate generalizability. A key limitation, however, is that the data collection setup assumes students are consistently facing a laptop, which restricts ecological validity; in real classroom settings, students often shift attention across multiple targets such as the blackboard, projector screens, and the lecturer.
\chapter{Analysis and Design}

This chapter details the theoretical analysis on which we proposed a novel BSN to enable measuring the engagement of individual students in a college STEM classroom while attending a lecture. This BSN, with the right set of algorithms deployed on it, can help mitigate the problem of high attrition rates among STEM majors by providing fast, objective feedback to professors and decision makers, enabling early intervention for students at risk. This analysis serves as the foundation upon which we design an effective BSN. The outcome of this chapter is the specification and requirements of the proposed BSN. This chapter is pivotal to this work: Chapter~4 presents an implementation of the BSN requirements developed here, while Chapter~5 describes the experiments conducted after implementation to evaluate the system’s performance and confirm that it satisfies the specifications and requirements established in this chapter.

\section{Analysis}

\subsection{Problem: High Attrition Rate Among Students Pursuing STEM Majors}

Studies show that, in the United States, for example, 50–60\% of students who enter college intending to complete a STEM degree ultimately either switch to a non-STEM field or leave college without earning a degree, despite being academically capable ~\cite{Chen2015, Chen2013, Olson2012}.

\subsection{Root Cause: Poor Performance in Foundational STEM Gateway Courses}

Poor performance in gateway STEM courses is a major factor contributing to student attrition from STEM pathways. These introductory, high-demand courses often determine whether students persist in a STEM major or choose to switch to a non-STEM field, and in 
many cases, they can even influence a student’s decision to leave college entirely \cite{Aulck2017, Chen2015}. Difficulties in mastering foundational material, combined with the pressure and pacing of these courses, create a critical barrier that disproportionately affects students who otherwise might succeed in STEM if provided timely support.

\subsection{Mitigation: Monitor Student Engagement During STEM Classes and Provide Support for Students at Risk}
A substantial body of research has demonstrated that higher levels of SE are linked to better academic performance, greater likelihood of degree completion, lower dropout rates, and fewer negative behaviors in educational environments \cite{Carini2006, FREDRICKS2015, Adnan2021}. Collectively, these findings highlight SE as a critical factor in reducing attrition and emphasize the need to understand how students engage within courses in order to enhance teaching and learning and to support timely interventions for students at risk. Indeed, engagement has even been described as “The holy grail of learning.” \cite{Sinatra2015}.

\subsection{Goal: An Effective System for Automatically Measuring Individual Student Engagement in Real Time in STEM Classrooms}

This work aims to propose a system that enables the \textbf{measurement} of individual college students' engagement \textbf{effectively} while attending a lecture in a \textbf{STEM classroom}, and does so in a manner that is both \textbf{ethically acceptable} and \textbf{technically viable}. To be precise, each term highlighted in bold in the previously stated aim must be clearly defined and operationalized.

\subsubsection{Measurement Framework}

A central challenge in designing a system capable of measuring SEt is determining how the abstract construct of engagement can be translated into measurable components. Following the guidance of \cite{Sinatra2015}, we structure this analysis around three fundamental questions:
\begin{enumerate}
    \item What is the \textit{construct definition} utilized?
    \item What is the \textit{grain size} for time, task, and agent?
    \item What are the \textit{indicators} to be collected?
\end{enumerate}

\paragraph{Construct Definition.}
We adopt the multidimensional framework of \cite{Fredricks2004}, which characterizes engagement in terms of behavioral, emotional, and cognitive components. These components are described in detail in Section~\ref{sec:engagement-definitions}. Importantly, the system cannot directly measure these dimensions; rather, it can capture only \emph{proxy indicators}. To map these proxy indicators to meaningful engagement estimates, complementary data—obtained via experience sampling methods (ESM) or systematic observation—must be collected to provide the ground truth for the measures generated by the proposed system.

\paragraph{Grain Size.}
Following \cite{Whitehill2014}, time is operationalized at a resolution of 10 seconds; the task is defined as attending a lecture in a STEM classroom; and the agent corresponds to the individual student being assessed. With this construct definition and grain size, our conceptualization aligns closely with that of Symonds \emph{et~al}.\ \cite{Symonds2024}.

The discussion of indicators is postponed until after the next subsection,  as the measurement methods—and therefore the indicators themselves—depend on the objectives established in that section.

\subsection{Effectiveness}
\label{sec:effectiveness}
By an effective system, we mean a system capable of mitigating the problem at hand. This necessitates that the system introduce minimal disruption to the normal flow of the lecture. Otherwise, the system may produce biased measurements if students notice that they are being monitored and consequently alter their behavior. Worse still, the system could contribute to additional disengagement if its presence in the classroom causes students to feel anxious or uncomfortable. Precisely for the system to be effective, it must fulfill five objectives:
\begin{multicols}{3}
\begin{enumerate}
    \item Non-intrusive
    \item Non-invasive
    \item Non-stigmatizing
    \item Real-time
    \item Automatic
\end{enumerate}
\end{multicols}
We acknowledge that the five objectives just mentioned above are somewhat broad and require greater precision. Therefore, each objective is defined more explicitly as follows.

A \textbf{Non-intrusive} system does not disrupt classroom activities, distract students, or alter their natural behavior. More concretely, a system is non-intrusive if it does not intercept the direct line of sight between the student and the professor, the projector, the blackboard, the student's notes, or the student's laptop. 

A \textbf{non-invasive} system requires no physical contact between itself and the student and avoids any procedure that interacts with the body in a medical sense. Concretely,  a system is non-invasive if the only physical contact with it occurs when a student voluntarily interacts with it at the beginning of class to provide consent and initiate the system; no sensors may be worn, attached, or applied to the student’s body at any point during operation.

A \textbf{non-stigmatizing} system ensures that no student is singled out, labeled, or made to feel judged as a result of the sensing process. Specifically, a system is non-stigmatizing if students are free to choose any available seat in the classroom, and it does not require designated seating or special placement for individual students. Furthermore, no student may be marked or tagged in any way during operation, as such practices would differentiate students from their peers and risk creating feelings of discomfort or stigma.

A system is called \textbf{real-time} if it captures, processes, and provides engagement-related information within a specified time window~$\Delta t$.

A system is called \textbf{automatic} if, once initialized, it requires no further human intervention during the operation.

\subsection{Engagement Indicators}
As detailed in Section~\ref{sec:eng_meas} and Section~\ref{sec:realtime_meas}, six methods are available in the literature to measure SE. However, not all of these methods align with the five objectives established in the previous subsection for the system to be effective. The goal of this discussion is to determine which methods, and therefore which indicators, are suitable for our system.

The \textit{student self-report surveys} and \textit{administrative data} methods do not fulfill the real-time objective and are therefore excluded from consideration. The \textit{teacher ratings} method satisfies neither the real-time nor the automatic objectives and is also unsuitable for our system.

The \textit{observational measures} method fulfills four of the stated objectives; however, it is not automatic. Despite this limitation, it remains a valuable tool for validating the results produced by an automatic, real-time system such as the one we aim to design. An automatic SE measurement system typically requires a labeled dataset (ground truth) for training. Because such data raises privacy concerns, no publicly available high-quality SE datasets exist for college STEM classrooms. This necessitates the collection of our own private dataset, and thus the system must support this capability. It is important to emphasize that dataset collection is a temporary step performed only once for system validation. In this context, observational methods can be used to manually annotate the collected data, which is acceptable since annotation is required only during development and not during deployment. Finally, because data collection is a serious matter, the system must ensure that all collection procedures adhere strictly to the regulations and policies of the University of Louisville Institutional Review Board.

The \textit{ESM} method does not satisfy the real-time, automatic, or non-intrusive objectives and therefore cannot be used in our system. However, similar to observational measures, it may still be employed to collect direct indicators from students immediately before or after class sessions to help validate the results generated by the automatic system.

This leaves us with the final method, \textit{real-time measures}. As the name suggests, this method supports measuring SE on a timescale ranging from a few seconds to a few minutes. Moreover, when combined with appropriate CV and ML algorithms and a high-quality dataset, it can be fully automated. Real-time methods support five types of indicators, as detailed in Section~\ref{sec:realtime_meas}; however, not all of them align with the five objectives stated previously. For example, HRV and EEG require invasive sensors and are therefore unsuitable. Fortunately, three of the five indicators—facial expressions, body actions, and eye gaze—can be collected using noninvasive sensors, such as a camera. The remaining two objectives, non-intrusive and non-stigmatizing, depend on the physical deployment of the system and will be proven later.

To conclude this section, the \textbf{real-time measures} method is the most suitable among the available approaches, as it aligns with the five objectives stated in the previous subsection. This method enables the indirect measurement of the three components 
of engagement using three different camera-based \textbf{biometric} indicators, as follows:
\begin{enumerate}
    \item \textbf{Behavioral engagement:} via body actions and movement patterns.
    \item \textbf{Emotional engagement:} via facial expressions and affective cues.
    \item \textbf{Cognitive engagement:} via eye movement patterns and gaze direction.
\end{enumerate}

\subsection{Ethics}
Ethical considerations are central to the design and deployment of any system intended to measure SE, particularly when it relies on biometric or video-based sensing. Such a system must respect student autonomy, protect privacy, and minimize any potential harm or discomfort arising from its presence in the classroom. This requires clear and transparent communication regarding what data are collected, how they are used, and who has access to them, as well as ensuring that participation is voluntary and informed. Moreover, data collection and storage procedures must adhere strictly to the regulations and policies of the Institutional Review Board (IRB), with safeguards in place to prevent misuse, unauthorized disclosure, or identification of individual students. Ultimately, the ethical foundation of the system must prioritize student well-being and trust, ensuring that the measurement process enhances learning rather than compromising the dignity or rights of those being observed.

\subsection{Environment}

The proposed system operates within a traditional college STEM classroom—an in-person lecture setting in which students are physically present, and an instructor delivers course material using a projector, whiteboard, or a combination of instructional media. We formalize this environment as the tuple
\[
(E, S, P),
\]
where $E$ denotes classroom calibration parameters, $S$ is the set of students, and $P(t)$ represents the professor’s feature vector at time~$t$. The student set is defined as
\[
S = \{\, s_i(t) \mid 1 \le i \le N_C \,\},
\]
where $s_i(t)$ is the feature vector describing student $i$ at time $t$, and $N_C$ is the maximum classroom capacity. Both $s_i(t)$ and $P(t)$ evolve over time, reflecting the dynamic and interactive nature of a live classroom environment.

The calibration parameters $E$ capture all fixed properties of the classroom relevant to system operation. These include the 3D geometry of the room, the locations and poses of instructional elements such as the blackboard and projector screen, and the placement and intrinsic parameters of fixed cameras used during development mode. In general, $E$ consists of any static quantities required by the BSN for accurate sensing, alignment, or spatial reasoning throughout a class session.

\subsection{Technical Viability}
For the proposed system to be technically viable, it must be feasible to deploy, operate, and maintain within the constraints of a typical college STEM classroom. This requires that the system rely on affordable, readily available hardware; run efficiently on resource-constrained edge devices; and integrate robust CV and ML algorithms capable of processing data in real time. The system must also tolerate variability in classroom conditions, including lighting, seating arrangements, and student movement, without significant degradation in performance. Furthermore, the system should be scalable, modular, and compatible with standard networking and data-management practices to support long-term deployment. In essence, technical viability ensures that the system can function reliably in real-world educational environments, not only in controlled laboratory settings.

Building on these insights, the next section translates these engagement indicators and objectives into concrete system requirements that guide the design of the proposed BSN.

\subsection{Sensing Technology}
From the analysis conducted in the previous subsections, we conclude that the sensing technology must be camera-based and that three types of indicators can be captured: facial expressions, eye gaze, and body actions. The next question, therefore, is which topology should be used to deploy these cameras. The answer to this question is addressed in the next subsection.

\subsection{Camera Deployment Topology}

Accurate capture of facial expressions and eye gaze requires high-resolution images of the student's face, which in turn necessitates positioning the camera close to the student’s face—typically on the student’s desk, facing the student alongside their laptop or tablet—to record subtle details. Such a camera can operate with a narrow field of view (FOV), and its pose can be freely adjusted to ensure that the student’s upper body, particularly the face, is consistently within view.

In contrast, capturing body actions requires cameras to be placed at a greater distance, facing students from above, so that the student’s full upper body remains within the FOV. These cameras can maintain a fixed position and are typically mounted on classroom walls or ceilings, with each camera covering multiple students within its field of view.

This distinction yields two possible deployment topologies: a \textit{dynamic} topology, in which each student is equipped with a device we call a student processing unit (SPU) having a narrow-FOV camera positioned freely across the classroom; and a \textit{static} topology, in which a fixed number of wide-FOV cameras are deployed per classroom. The question, then, is which topology is most suitable for our system. To address this, we now examine the advantages and disadvantages of each approach.
\subsubsection{Static Topology}

The static topology offers advantages in scalability and cost. A single classroom, including those with large capacities, can be covered using only a few cameras. Moreover, this approach does not require specialized hardware; any off-the-shelf IP camera with a suitable field of view and resolution can be used.

However, this topology also presents several drawbacks. First, it does not satisfy the \emph{non-stigmatizing} objective. For example, in a classroom where some students prefer not to be monitored or recorded, it becomes nearly impossible to honor their preferences without restricting their freedom to choose their seats. Second, because the cameras are positioned far from students’ faces, they cannot capture high-quality facial imagery and therefore cannot reliably extract facial expressions or eye gaze. Third, the topology raises privacy concerns since images must be transmitted to a server for processing rather than processed at the edge. Fourth, while camera installation requires only a one-time effort, the fixed placement may affect classroom usability for other courses, as students unaffiliated with the monitoring system may feel uncomfortable due to the camera presence.

Despite these limitations, the static topology is well-suited for collecting an SE dataset that can later be annotated using observational protocols.

\subsubsection{Dynamic Topology}

The dynamic topology offers several advantages. First and foremost, it satisfies the \emph{non-stigmatizing} objective, since each student is assigned a dedicated SPU and is therefore free to choose any seat in the classroom. Second, it mitigates privacy concerns because all processing occurs on the edge; no student-identifiable data, such as images or videos, is transmitted to a centralized server for further processing. Furthermore, if a student feels uncomfortable for any reason, the SPU can be turned off or removed entirely. Third, this topology reliably captures two of the engagement indicators, namely facial expressions and eye gaze. Finally, once the class session ends, the SPUs can be collected quickly and stored securely, ensuring that the classroom can be used normally for other courses without affecting students not involved in the monitoring process.

Despite its versatility, the dynamic topology also has disadvantages. The system cost scales linearly with classroom size, since each student requires an SPU. This is not a major issue for small and medium-sized classrooms, which are typical in STEM courses. Additionally, the topology requires designing a custom unit—i.e., the SPU. However, this is a one-time effort and can be accomplished using readily available edge-computing devices such as the Raspberry Pi\cite{RaspberryPiLtd2025} or NVIDIA Jetson\cite{jetson_orin_nano}. A further limitation is that the SPU has restricted computational capability, which may prevent it from running computationally heavy algorithms. This challenge can be mitigated by selecting efficient algorithms or employing techniques such as quantization and pruning to reduce computational cost. Finally, the dynamic topology cannot reliably capture students’ body actions.

In summary, the dynamic topology is the only configuration that satisfies the \emph{non-stigmatizing} objective and is therefore the topology of choice for deployment. However, during the development phase, the static topology is necessary for capturing and recording our private SE dataset, which will later be annotated by trained coders and used to validate and benchmark the proposed system.

\subsection{The Biometric Sensor Network}

We are now ready to introduce the concept of a \textit{Biometric Sensor Network} (BSN), which consists of two types of sensing nodes: \textit{dynamic nodes}, called Student Processing Units (SPUs), and \textit{static nodes}, called Classroom Fixed Units (CFUs). Both node types communicate with a central server over a network to coordinate sensing, processing, and data management. To ensure that the BSN satisfies the remaining objective of being \emph{non-intrusive}, the network technology must be wireless, and each SPU must be battery powered. This prevents cable clutter in the classroom and allows the SPU to be placed or repositioned freely, preserving the natural classroom environment and students’ freedom of movement. In addition, the BSN includes a front-end client in the form of a web-based dashboard, which provides a unified interface to control, monitor, and operate the entire BSN during classroom sessions.

In addition, the BSN supports two modes of operation: a \emph{development mode} and a \emph{deployment mode}. In development mode, both SPUs and CFUs are used to capture raw video of students for the purpose of constructing a private SE dataset. The CFU footage is later annotated by trained coders to produce engagement labels for each 10-second segment for every student. In deployment mode, CFUs are no longer used. Instead, each SPU runs the necessary algorithms locally to infer SE from 10-second video segments captured by its own camera, and then transmits the inferred SE values to the central server for storage and real-time monitoring.

\section{System Requirements}
\label{sec:sys_req}
Having established the structure, operational modes, and functional objectives of the proposed BSN, we now turn to the system requirements and design considerations necessary to realize this network in practice. The next subsections detail the hardware and software components of the SPUs and CFUs, the communication architecture, and the processing pipeline that enables real-time engagement inference within classroom environments.

\subsection{SPU Hardware Requirements}

The Student Processing Unit (SPU) serves as the core sensing and computation node in the dynamic topology of the BSN, and its hardware requirements are defined to ensure that it can operate reliably, securely, and in compliance with the non-intrusive, non-invasive, and non-stigmatizing design objectives. Each SPU must integrate a high-quality camera capable of capturing facial expressions and eye gaze, a compute module capable of running lightweight CV/ML inference models in real time, and a touchscreen interface through which students can provide consent at the beginning of each session. To preserve classroom flexibility and avoid cable clutter, the SPU must operate wirelessly and be powered by a battery system that provides safe and uninterrupted operation throughout the class period. In addition, precise clock synchronization is required to align inference windows across devices, and encrypted temporary storage is necessary to support secure buffering before data transmission.

The enclosure plays an essential role in both usability and security. It must be compact, non-distracting, and thermally well-ventilated, while also ensuring strong physical security. In particular, external ports such as USB or Ethernet must be physically hidden, covered, or internally routed so that students cannot access them during operation. This prevents unauthorized data extraction, device tampering, or circumvention of system security controls. Collectively, these requirements ensure that the SPU functions as an effective, secure, and ethically responsible sensing device within the proposed BSN. The complete set of hardware requirements for the SPU is summarized in Table~\ref{tab:SPU_HW}.

\begin{longtable}{|p{4cm}|p{10cm}|}
\caption{SPU Hardware Requirements}
\label{tab:SPU_HW} \\
\hline
\textbf{Component} & \textbf{Requirements} \\
\hline
\endfirsthead
\hline
\textbf{Component} & \textbf{Requirements} \\
\hline
\endhead
\hline
\endfoot
\hline
\endlastfoot
Camera Module & High-resolution 720p+ camera, narrow FOV, low-light capable, minimum 15 fps.
\\ \hline
Compute Unit & ARM-based single board computer (SBC) (e.g., Raspberry Pi 5 \cite{RaspberryPiLtd2025}) with optional GPU support for real-time CV/ML inference.
\\ \hline
Touchscreen Interface & Integrated touch display for collecting student consent before each session, displaying system status, and enabling basic interaction and control. The touchscreen must automatically turn off within 30 seconds after consent is provided to minimize distraction, and it must be reactivatable by touch to allow the student to end the recording session or power off the device if desired.
\\ \hline

Battery System & Minimum 2 hours of continuous use; Maximum charging time is 2 Hours; Overcharge, discharge, overcurrent, and thermal protection.
\\ \hline
Network Interface & Wireless interface (Wi-Fi 5/6) for reliable, low-latency communication with the central server.
\\ \hline
Clock Sync. & NTP-based time alignment with jitter less than 500 ms.
\\ \hline
Enclosure & Compact, non-obstructive housing with proper thermal ventilation; no distracting lights or visible indicators; the enclosure must mechanically support the full weight of the SPU and be designed with appropriate balance so it can stand stably on a desk without additional mounts, adhesives, supports, or screws.
\\ \hline

Physical Security & All external ports (e.g., USB, Ethernet) must be physically hidden or rendered inaccessible during operation, exposing only a power button and a protected battery-charging port to prevent tampering or unauthorized data access.
\\ \hline
Local Storage & Supports at least 30 GB of encrypted temporary buffering; automatic deletion after processing and upload.
\\ \hline

\end{longtable}

\subsection{CFU Hardware Requirements}

The Classroom Fixed Unit (CFU) serves as the static sensing node within the BSN and is used primarily during the development phase to support large-scale data collection and annotation. Unlike the SPU, which operates in proximity to individual students, the CFU is positioned at fixed locations in the classroom—typically overhead—to capture wide-angle views of multiple students simultaneously. Its primary purpose is to enable the construction of a labeled SE dataset through synchronized video capture that can later be annotated by trained coders. Consequently, the hardware requirements for the CFU emphasize high-resolution wide-field imaging, stable physical installation, reliable network connectivity, and precise time synchronization. These requirements are summarized in Table~\ref{tab:CFU_HW} and ensure that the CFU can operate continuously, accurately, and securely during classroom recording sessions.

\begin{longtable}{|p{4cm}|p{10cm}|}
\caption{CFU Hardware Requirements}
\label{tab:CFU_HW} \\
\hline
\textbf{Component} & \textbf{Requirements} \\
\hline
\endfirsthead
\hline
\textbf{Component} & \textbf{Requirements} \\
\hline
\endhead
\hline
\endfoot
\hline
\endlastfoot

Wide-FOV Camera & 92–120° FOV, overhead mounted; $4$K resolution.
\\ \hline
Mounting System & Ceiling/wall fixed mount.
\\ \hline
Network Interface & Wired (10 Gbs Ethernet) or wireless (Wi-Fi 5/6); consistent streaming performance.
\\ \hline
Power Supply & Permanent AC power or Power Over Ethernet (PoE).
\\ \hline
Clock Sync. & NTP-based time alignment with jitter less than 500 ms.
\\ \hline
\end{longtable}

\subsection{SPU Software Requirements}

The software stack running on each SPU is critical to ensuring that the device operates securely, autonomously, and in real time, in accordance with the system objectives established in the earlier analysis. Because the SPU performs all engagement inference locally—without transmitting raw video—the software must support efficient on-device CV/ML processing, secure communication protocols, and reliable synchronization with the central server. In addition, the software must enforce strict security guarantees, including a hardened operating system, encrypted data handling, and authenticated over-the-air updates, to prevent tampering and protect student privacy. Robust power management and fault-tolerant networking further ensure continuous operation throughout a class session. Table~\ref{tab:SPU_SW} summarizes the key software requirements necessary for the SPU to function as an effective, secure, and privacy-preserving sensing node within the BSN.

\begin{longtable}{|p{4cm}|p{10cm}|}
\caption{SPU Software Requirements}
\label{tab:SPU_SW} \\
\hline
\textbf{Component} & \textbf{Requirement} \\
\hline
\endfirsthead
\hline
\textbf{Component} & \textbf{Requirement} \\
\hline
\endhead
\hline
\endfoot
\hline
\endlastfoot

Operating System & Hardened 64-bit Linux distribution with minimized attack surface and disabled unnecessary services. \\ \hline

CV/ML Inference Engine & Real-time models for face detection, gaze estimation, and behavioral analysis optimized for edge devices. \\ \hline

On-device Processing & All inference is performed locally; raw video must never be stored or transmitted. \\ \hline

Time Synchronization & NTP-based synchronization with maximum jitter of 500 ms to align 10-second inference windows. \\ \hline

Communication & TLS 1.3 encrypted communication with automatic retry, reconnection, and failover handling. \\ \hline

Power Manager & Battery monitoring, low-power alerts, and controlled safe-shutdown logic. \\ \hline

Wi-Fi Registration & Software must support onboarding SPUs onto the UofL Wi-Fi network using approved campus authentication protocols. \\ \hline

Node Registration & SPUs must obtain a signed SSL certificate from the server via a one-time pre-shared key to enable authenticated operation in designated classrooms. \\ \hline

Automatic Session Control & The software must automatically detect active class sessions, enter session mode at start time, and exit session mode upon session completion. \\ \hline

Consent Gatekeeping & The SPU must not activate its camera unless the student provides explicit consent through the touchscreen interface at the beginning of each session. \\ \hline

System Status Bar & The GUI must display clock, Wi-Fi status, certificate validity, secure-connection status, and battery level. \\ \hline

System Power Controls & The GUI must provide a safe interface for shutdown and reboot operations. \\ \hline

OTA Updates & Support for authenticated, signed over-the-air updates for both software and firmware. \\ \hline

\end{longtable}

\subsection{Security and Data Protection for Edge Nodes}

The edge devices in the BSN, namely the SPUs and CFUs, must enforce strong security and privacy protections, as they operate in close physical proximity to students and handle potentially sensitive visual data. Security at the edge focuses on minimizing the attack surface, ensuring that devices cannot be tampered with, and guaranteeing that no raw video or identifying biometric information ever leaves the device. To achieve this, the edge nodes must implement encrypted storage, authenticated boot mechanisms, strong device-level authentication, and strict access control. In addition, all processing of engagement indicators must occur locally to preserve privacy, with only nonidentifying features transmitted to the server. These requirements ensure that edge devices remain secure, privacy-preserving, and compliant with institutional regulations. The key security and data-protection requirements for SPUs and CFUs are summarized in Table~\ref{tab:Edge_Security}.

\begin{longtable}{|p{4cm}|p{10cm}|}
\caption{Security and Data Protection Requirements (Edge Devices: SPUs/CFUs)}
\label{tab:Edge_Security} \\
\hline
\textbf{Aspect} & \textbf{Requirement} \\
\hline
\endfirsthead

\hline
\textbf{Aspect} & \textbf{Requirement} \\
\hline
\endhead

\hline
\endfoot

\hline
\endlastfoot

Encryption & Local storage must use AES-256 encryption; raw video must never be transmitted off-device. \\ \hline

Secure Boot & Device must verify firmware integrity at boot to prevent unauthorized modifications. \\ \hline

Device Authentication & Mutual TLS authentication using hardware-backed certificates. \\ \hline

Data Minimization & Raw video must be deleted immediately after inference; only nonidentifying indicators may be sent to the server. \\ \hline

Access Control & SSH must be disabled or restricted to key-based access; no default or shared credentials. \\ \hline

Privacy by Design & Only behavioral indicators may be extracted; no biometric identification or student re-identification allowed. \\ \hline

\end{longtable}

\subsection{Server Software Requirements}

The central server forms the core coordination and data-management layer of the BSN. It must provide secure and reliable services for authenticating users and devices, ingesting engagement data from SPUs, coordinating active classroom sessions, and supporting real-time monitoring through a web-based dashboard. Because SPUs operate autonomously at the edge, the server is responsible for validating node identity, authorizing access, aggregating engagement estimates, and maintaining system-wide consistency through periodic health checks. In addition, the server must implement robust analytics capabilities, enforce role-based access control, and maintain comprehensive audit logs to support debugging, anomaly detection, and institutional compliance. The key software requirements for the server are summarized in Table~\ref{tab:Server_SW}.

\begin{longtable}{|p{4cm}|p{10cm}|}
\caption{Server Software Requirements}
\label{tab:Server_SW} \\
\hline
\textbf{Component} & \textbf{Requirement} \\
\hline
\endfirsthead
\hline
\textbf{Component} & \textbf{Requirement} \\
\hline
\endhead
\hline
\endfoot
\hline
\endlastfoot

Backend API & Provides APIs for user registration and authorization, node registration and 
authorization, periodic health checks, and secure ingestion of engagement data.
\\ \hline
Database & Stores engagement time-series data with AES-256 encryption and automated backup rotation.
\\ \hline
Dashboard Interface & Web interface for system monitoring, configuration, and real-time engagement visualization.
\\ \hline
Analytics Engine & Computes aggregated engagement metrics per student, class session, and time window.
\\ \hline
User Management & Enforces role-based access control (e.g., admin, instructor, researcher).
\\ \hline
Logging and Auditing & Maintains comprehensive logs for debugging, anomaly detection, and regulatory compliance.
\\ \hline
\end{longtable}

\subsection{Security and Data Protection for Server}

The server infrastructure of the BSN must enforce strong security and privacy guarantees, as it serves as the central point for data aggregation, device identity validation, and long-term storage of engagement estimates. In addition to providing encrypted transport channels and secure storage for engagement data, the server must support an on-premises, access-controlled object storage solution for dataset archival and operate its own certificate authority to issue and manage SPU certificates, including revocation when necessary. Robust identity and access management—with MFA and role-based access control—is required to restrict system operations to authorized personnel. Furthermore, the server must support intrusion detection and incident response mechanisms and provide the ability to isolate compromised nodes to preserve overall system integrity. All security controls must comply with institutional and regulatory requirements such as IRB and FERPA. The principal security and data-protection safeguards implemented at the server level are summarized in Table~\ref{tab:Server_Security}.

\begin{longtable}{|p{4cm}|p{10cm}|}
\caption{Security and Data Protection Requirements (Server)}
\label{tab:Server_Security} \\
\hline
\textbf{Aspect} & \textbf{Requirement} \\
\hline
\endfirsthead

\hline
\textbf{Aspect} & \textbf{Requirement} \\
\hline
\endhead

\hline
\endfoot

\hline
\endlastfoot

Transport Security & All device–server and user–server communications must use TLS 1.3.
\\ \hline
Database Protection & Engagement data stored at rest must use AES-256 encryption with automated secure backup rotation.
\\ \hline
Identity \& Access Management & System must enforce MFA and role-based access control (RBAC) following least-privilege principles.
\\ \hline
Secure Object Storage & Supports on-premises, access-controlled S3-compatible storage (e.g., MinIO) for dataset collection and archival.
\\ \hline
Certificate Authority (CA) & Issues and manages SPU certificates through an on-premises CA to support authentication and certificate revocation.
\\ \hline
Incident Response & Server must support intrusion detection/prevention (IDS/IPS) and isolate or revoke compromised SPUs.
\\ \hline
Compliance & All processing and storage must comply with IRB, FERPA, and institutional data-protection policies.
\\ \hline

\end{longtable}

\section{Summary}

\begin{sidewaysfigure}
    \centering
    \includegraphics[width=0.95\textheight]{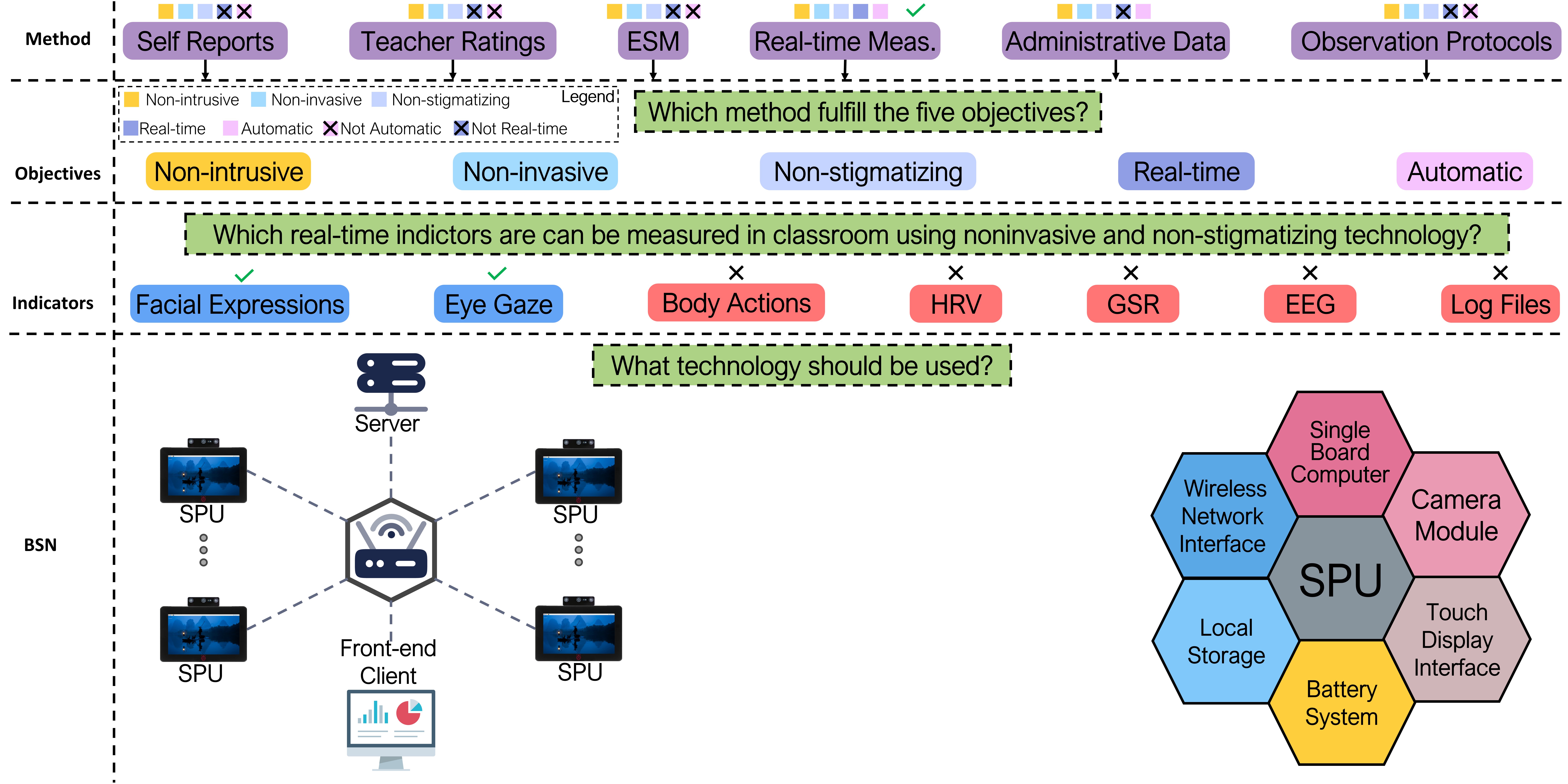}
    \caption{Design-decision workflow summarizing the selection of SE measurement methods, real-time indicators, and sensing technology leading to the final SPU-based BSN architecture.}
    \label{fig:DES_DEC_DIAG}
\end{sidewaysfigure}

Figure~\ref{fig:DES_DEC_DIAG} summarizes the design process that led to the adoption of an SPU-based BSN. Beginning with six established methods for measuring SE, each was evaluated against the five required system objectives: non-intrusive, non-invasive, non-stigmatizing, real-time, and automatic. Only the real-time measurement approach satisfies these criteria. Within this method, a further evaluation of available indicators showed that facial expressions, eye gaze, and body actions are the only signals that can be captured using non-invasive and non-stigmatizing technology suitable for a live classroom environment. These constraints naturally lead to a camera-based system, which must be portable, wireless, battery-operated, and capable of on-device processing. The resulting design decision converges on the SPU architecture, which integrates a single-board computer, camera module, wireless interface, local storage, battery subsystem, and touch display for consent and control. 

\chapter{BSN Implementation}
This chapter presents the complete hardware and software implementation of the proposed BSN, translating the system requirements derived in the previous chapter into a functional, deployable platform. Whereas the earlier analysis established \textit{what} the system must achieve and \textit{why} each design choice is necessary, this chapter describes \textit{how} the system is realized in practice. The discussion is organized around the two principal components of the BSN: the Student Processing Units (SPUs), which operate as portable edge nodes responsible for real-time engagement inference, and the central server, which coordinates device authentication, session management, data aggregation, and visualization.

On the hardware side, the chapter details the SPU system architecture, component selection, integration, and configuration. On the software side, we describe the development of the SPU software stack for on-device CV/ML inference, consent collection, network communication, and secure operation, as well as the 
server-side backend responsible for data ingestion, analytics, access control, and system 
monitoring.

Together, the hardware and software implementations presented in this chapter constitute a fully functional BSN capable of operating in both development and deployment modes, supporting dataset collection, real-time engagement measurement, and secure classroom integration. This chapter provides all engineering details necessary to reproduce, evaluate, and extend the proposed system.
\begin{wrapfigure}{r}{0.50\textwidth}
    \centering
    \vspace{40pt}
    \includegraphics[width=0.48\textwidth]{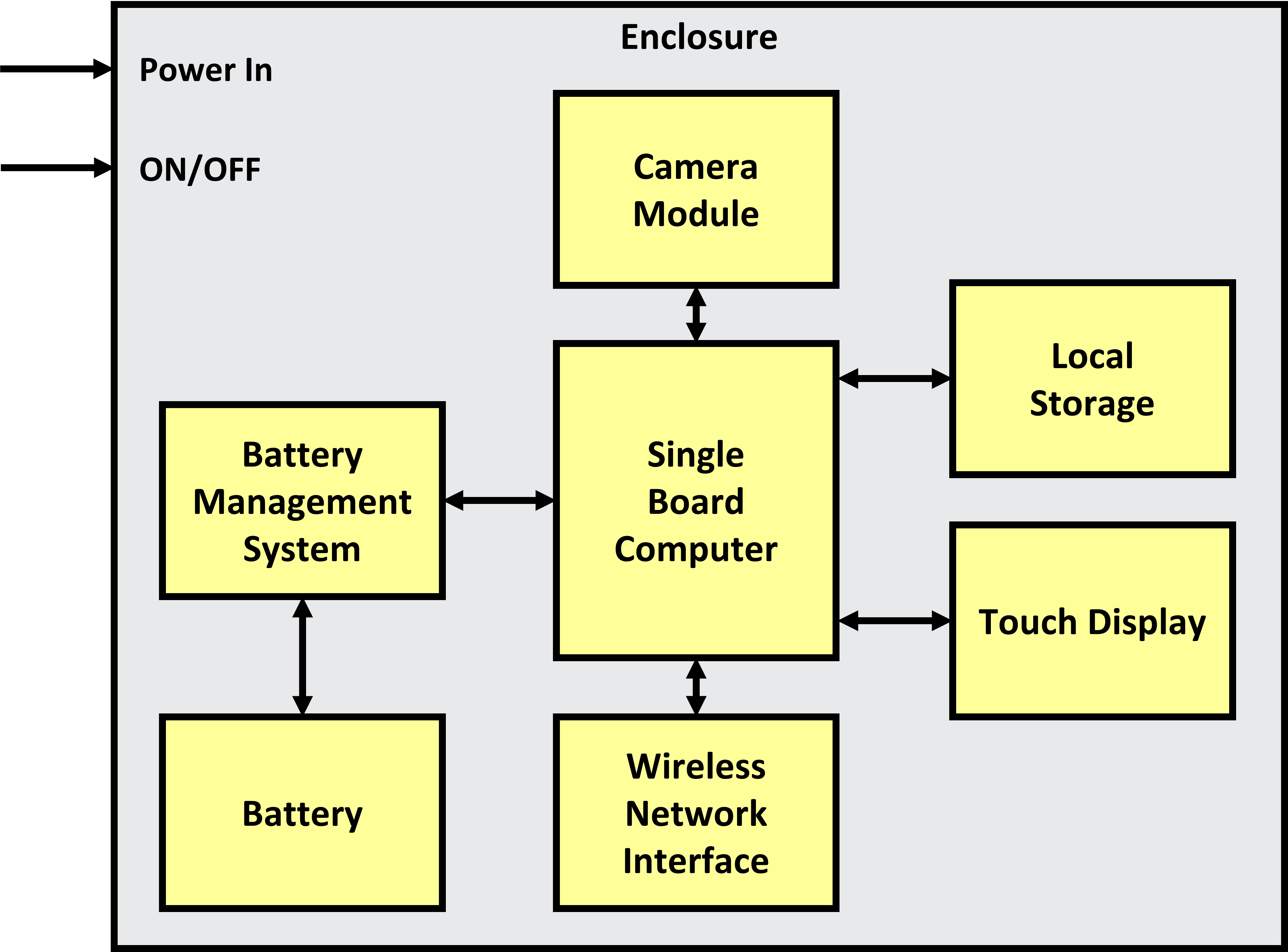}
    \caption{System architecture of the SPU hardware, illustrating the interaction between the SBC, camera module, touch display, wireless network interface, battery management subsystem, and local storage.}
    \label{fig:SPU_HW_ARCH}
\end{wrapfigure}

\section{SPU Hardware}

As established in the previous chapter, the SPU hardware must integrate six essential functional components: a single-board computer (SBC), a camera module, a touch display, a wireless communication interface, a battery management subsystem, and a local storage unit. Figure~\ref{fig:SPU_HW_ARCH} illustrates the overall architecture of the SPU and the interactions between these components. In this section, we identify suitable off-the-shelf hardware elements that fulfill these requirements and provide a practical realization of the proposed design.

\subsection{SBC}
A key design decision concerns the choice of a single-board computer (SBC), as it directly affects the computational throughput, latency performance, power consumption, physical footprint, and available peripheral interfaces. Several commercially available SBCs were evaluated against these criteria, and a summary of their characteristics is provided in Table~\ref{tab:SBC_comparison}. The comparison accounts for processing capability, device dimensions, cost, typical power usage, and availability of required interfaces such as SD card storage, USB3, Wi-Fi, DSI, and I\textsuperscript{2}C. Based on this evaluation, the Raspberry~Pi~5\cite{RaspberryPiLtd2025} was selected because it offers an optimal balance between performance, energy efficiency, cost, and ecosystem maturity for the SPU’s operational setting. In addition, the Raspberry~Pi~5\cite{RaspberryPiLtd2025} simplifies the overall hardware design by integrating both a Wi-Fi~6 wireless interface and a native microSD storage slot, thereby eliminating the need for external networking modules and removable storage hardware.

\begin{table}[h!]
\centering
\scriptsize
\renewcommand{\arraystretch}{1.3}

\resizebox{\textwidth}{!}{%
\begin{tabular}{|l|p{4.2cm}|l|l|l|p{3cm}|}
\hline
\textbf{SBC} & \textbf{Compute Power} & \textbf{Size$^\ast$ (w×l×h)} & 
\textbf{Price} & \textbf{Power$^\ast$} & \textbf{Peripherals} \\
\hline

Raspberry Pi 5 \cite{RaspberryPiLtd2025} &
4× Cortex-A76 @ 2.4GHz  
\newline 20–25 GFLOPS &
85×56×18 mm &
\$45–145 &
2.5–12W &
\textbf{SD Card}, \textbf{USB3}, \textbf{WiFi 6}, \textbf{DSI}, \textbf{I2C} \\
\hline

Raspberry Pi 4B \cite{raspberrypi4b} &
4× Cortex-A72 @ 1.8GHz  
\newline 10–15 GFLOPS &
85×56×17 mm &
\$35–85 &
2.7–8W &
\textbf{SD Card}, \textbf{USB3}, \textbf{WiFi 5}, \textbf{DSI}, \textbf{I2C} \\
\hline

Jetson Orin Nano \cite{jetson_orin_nano} &
6-core ARM + Ampere GPU  
\newline 40–67 TOPS &
100×80×30 mm &
\$249 &
7–25W &
\textbf{SD Card}, \textbf{USB3}, \textbf{WiFi 6}, \textbf{DSI}, \textbf{I2C} \\
\hline

Orange Pi 5 Plus \cite{orangepi5plus} &
RK3588 (4×A76 + 4×A55)  
\newline 0.8 TFLOPS + NPU 6 TOPS &
100×64×18 mm &
\$130–220 &
5–15W &
\textbf{SD Card}, \textbf{USB3}, \textbf{DSI}, \textbf{I2C} \\
\hline

ROCK 5B \cite{rock5b} &
RK3588 (4×A76 + 4×A55)  
\newline 0.8 TFLOPS + NPU 6 TOPS &
100×72×18 mm &
\$150–220 &
3–15W &
\textbf{SD Card}, \textbf{USB3}, \textbf{DSI}, \textbf{I2C} \\
\hline

BeagleBone Black \cite{beagleboneblack} &
Cortex-A8 @ 1GHz  
\newline ~0.5 GFLOPS &
86×53×17 mm &
\$70 &
1–2W &
\textbf{SD Card}, \textbf{I2C} \\
\hline

BeaglePlay \cite{beagleplay} &
4× Cortex-A53 + PRUs  
\newline ~2 GFLOPS &
100×100×20 mm &
\$100–120 &
2–5W &
\textbf{SD Card}, \textbf{USB3}, \textbf{WiFi 5}, \textbf{I2C} \\
\hline

\end{tabular}
} 

\caption{Comparison of candidate single-board computers (SBCs) evaluated for the SPU. 
Metrics include compute performance, physical dimensions, cost, power requirements, and 
required peripherals.}
\label{tab:SBC_comparison}

\vspace{6pt}
\noindent\begin{minipage}{\textwidth}
{\footnotesize $^\ast$\textit{Size and power consumption values may vary depending on cooling 
accessories (e.g., heatsinks, fans), workload intensity, and vendor revisions. Values should 
be interpreted as typical operating ranges rather than strict guarantees.}}
\end{minipage}
\end{table}

\subsection{Camera Module}
The next design decision concerns the selection of the camera module to be integrated into the SPU. The Raspberry~Pi~5\cite{RaspberryPiLtd2025} supports two primary camera interfaces: the built-in CSI (Serial Camera Interface) for dedicated camera modules and the USB~3.0 port for higher-bandwidth external vision systems. Both categories offer viable options, and the choice depends on the required resolution, field of view, low-light performance, and the computational demands of real-time engagement inference.

While standard Raspberry Pi camera modules provide high-quality imaging with low power consumption and tight hardware integration, they rely entirely on the SBC for image processing. In contrast, modern smart camera modules incorporate on-board accelerators that can execute CV/ML workloads, thereby reducing the computational burden on the Pi~5. This is especially important for running face detection, gaze estimation, and expression recognition in real time.

After evaluating available options, the DepthAI OAK-D Lite~\cite{OAKDLite} camera module was selected as the preferred choice. It features a 4K RGB sensor capable of capturing 4K video at 30 frames per second (fps) and Full HD (FHD) video at 60~fps. In addition, it includes a mono-stereo depth pair operating at a resolution of 480p with frame rates of up to 120~fps, along with an integrated Intel Myriad~X\cite{Intel_MyriadX_4GB} AI accelerator providing up to 4~TOPS (approximately 1.4~TOPS usable for CV/ML workloads). This onboard processing capability substantially offsets the Raspberry~Pi’s limited hardware acceleration and enables efficient real-time inference on the SPU. Furthermore, the device is powered directly from the Pi’s USB port, simplifying the SPU hardware design by eliminating the need for additional power rails or external converters. The OAK-D Lite is also considerably more 
cost-effective than other depth-enabled, accelerator-equipped camera modules, making it an ideal choice for large-scale classroom deployment across SPU nodes. A comparison of possible camera options is provided in Table~\ref{tab:Camera_comparison}.

\begin{table}[h!]
\centering
\scriptsize
\renewcommand{\arraystretch}{1.3}

\resizebox{\textwidth}{!}{%
\begin{tabular}{|l|l|p{2.2cm}|p{2.2cm}|l|l|l|}
\hline
\textbf{Camera Module} &
\textbf{Interface} &
\textbf{Stereo Res. / FPS} &
\textbf{RGB Res. / FPS} &
\textbf{AI Acc.} &
\textbf{Power (W)} &
\textbf{Cost (\$)} \\
\hline

Pi Camera Module 3~\cite{PiCamera3} &
CSI-2 &
No &
12MP @ 30 fps &
No &
N/A &
30 \\
\hline

\textbf{OAK-D Lite}~\cite{OAKDLite} &
USB 3.0 &
480p @ 120 fps &
4K @ 30 fps &
1.4 TOPS &
5--7.5 &
149 \\
\hline

OAK-D S2~\cite{OAKDS2} &
USB 3.0 &
1MP @ 120 fps &
4K @ 30 fps &
1.4 TOPS &
5--7.5 &
299 \\
\hline

OAK-D Pro~\cite{OAKDPro} &
USB 3.0 &
1MP @ 120 fps &
4K @ 30 fps &
1.4 TOPS &
5--7.5 &
399 \\
\hline

RealSense D435x~\cite{RealSenseD435x} &
USB 3.0 &
FHD @ 90 fps &
FHD @ 90 fps &
No &
0.04--2.6 &
314--354 \\
\hline

\end{tabular}
} 

\caption{Comparison of camera modules compatible with the Raspberry~Pi~5, including interface type, stereo and RGB resolution with maximum frame rates, on-board AI acceleration, \textbf{power consumption}, and approximate cost.}
\label{tab:Camera_comparison}
\end{table}

\subsection{Local Storage and Touch Display}

The SPU requires sufficient local storage to ensure reliable operation and to buffer data during classroom sessions. A 64 GB Class~10 SD card is used, allocating approximately 10 GB for the operating system and software stack, with the remaining 56 GB reserved for temporary backup storage. This capacity allows the system to record OAK-D camera streams for at least two hours—well above the typical one-hour lecture duration—providing a safety margin in the event of network congestion or bandwidth limitations. Local buffering ensures that no engagement-relevant data is lost during transmission delays, and recordings can be automatically uploaded or deleted depending on the BSN’s mode of operation.

For the touch interface, the SPU uses the Raspberry Pi Touch Display~2 \cite{PiDisplay2}, which provides a 24-bit RGB panel with a native resolution of $720\times1280$ pixels. The display consumes approximately 2.5 W at maximum brightness, making it suitable for battery-powered portable operation. It is used primarily for consent collection, system status display, and basic user interaction, and automatically turns off during sessions to minimize distraction and reduce power consumption.

\subsection{Battery Management System}

Given the estimated peak power consumption of the SPU, approximately 12 W for the Raspberry Pi 5, up to 7.5 W for the OAK-D Lite camera, and around 2.5 W for the touch display, the total worst-case load is about 22 W. Moreover, the Raspberry Pi 5 power specification requires that its supply be capable of delivering 5 V at 5 A, which places additional constraints on the power subsystem. The battery management system must therefore satisfy two primary requirements: (i) provide a stable 5 V output with enough current headroom (up to 5 A) to reliably power the Pi 5 and its peripherals, and (ii) supply sufficient battery capacity to support at least one full class session without interruption.

To meet these requirements, we employ the Geekworm X1200 UPS HAT \cite{X1200}, which is specifically designed for the Raspberry~Pi~5 and supports two 18650 lithium-ion cells while providing a regulated 5.1~V output at up to 5~A continuous current. The X1200 interfaces with the Pi through pogo-pin contacts, eliminating the need for bulky connectors and reducing mechanical wear during assembly. The module also integrates over-current, over-discharge, and over-charge protection circuitry, making it a robust and reliable power solution for the SPU. Furthermore, the X1200 offers uninterrupted power capability and includes a fuel-gauge interface accessible over I\textsuperscript{2}C to enable safe shutdown signaling, which is essential for preventing filesystem corruption and ensuring reliable SPU operation in classroom deployments.

For this design, two 3500 mAh cells are used, yielding a nominal energy capacity of
\[
E_{\text{bat}} \approx 2 \times 3.7 \text{V} \times 3.5 \text{Ah} \approx 25.9 \text{Wh}.
\]
Assuming a conservative 90\% power-conversion efficiency, the usable energy delivered at the 
5 V rail is approximately $0.9 \times 25.9 \approx 23$ Wh. Under a continuous worst-case 
load of about 22 W, which corresponds to slightly more than one hour of operation.

In practice, however, the average power draw is considerably lower: the display is blanked during most of the session, the OAK-D Lite typically consumes around 5 W, and the Raspberry Pi 5 rarely operates at its peak power, drawing less than 5 W under moderate workloads. This brings the total steady-state consumption to under 12 W, which allows the SPU to operate for at least two hours on the available battery capacity.

\begin{wrapfigure}{r}{0.50\textwidth}
    \centering
    \vspace{0pt}
    \includegraphics[width=0.5\textwidth]{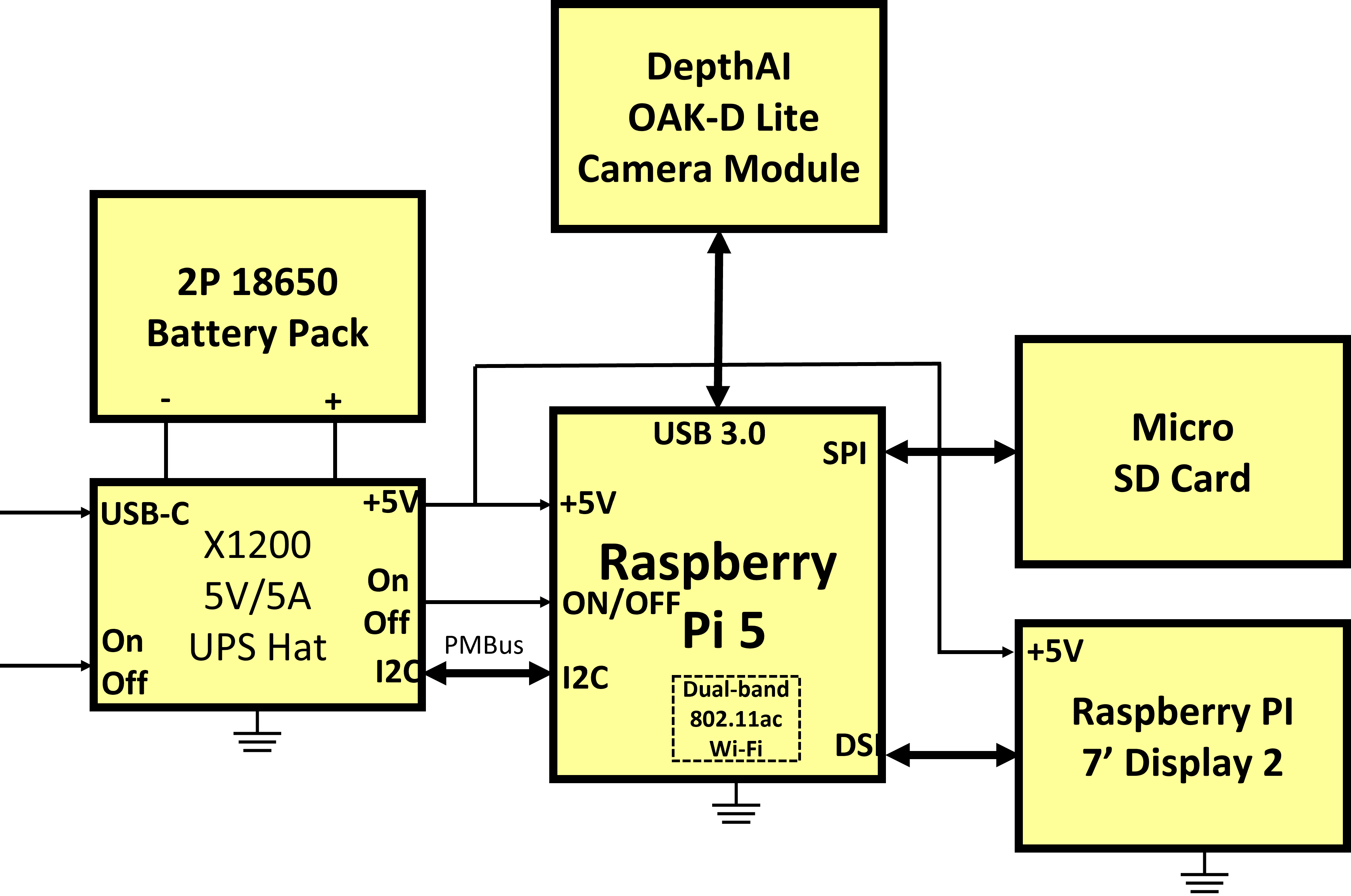}
    \caption{System-level schematic diagram of the SPU, showing the electronic components and 
    their interconnections.}
    \label{fig:SPU_HW_SCH}
\end{wrapfigure}

Figure~\ref{fig:SPU_HW_SCH} presents the complete electronic system schematic of the SPU, illustrating the interconnections among all selected components, including the Raspberry Pi 5, OAK-D Lite camera, battery management subsystem (X1200), dual 18650 battery pack, local storage, and touch display. This schematic consolidates the hardware decisions discussed in the previous subsections and serves as the blueprint for the SPU’s electrical integration and subsequent PCB and enclosure design.

\subsection*{SPU Enclosure\footnote{The enclosure was designed by a contracted freelance mechanical designer based on system specifications provided by the author.}}
\addcontentsline{toc}{subsection}{SPU Enclosure}

\begin{figure}[h!]
    \centering
    \includegraphics[width=0.85\textwidth]{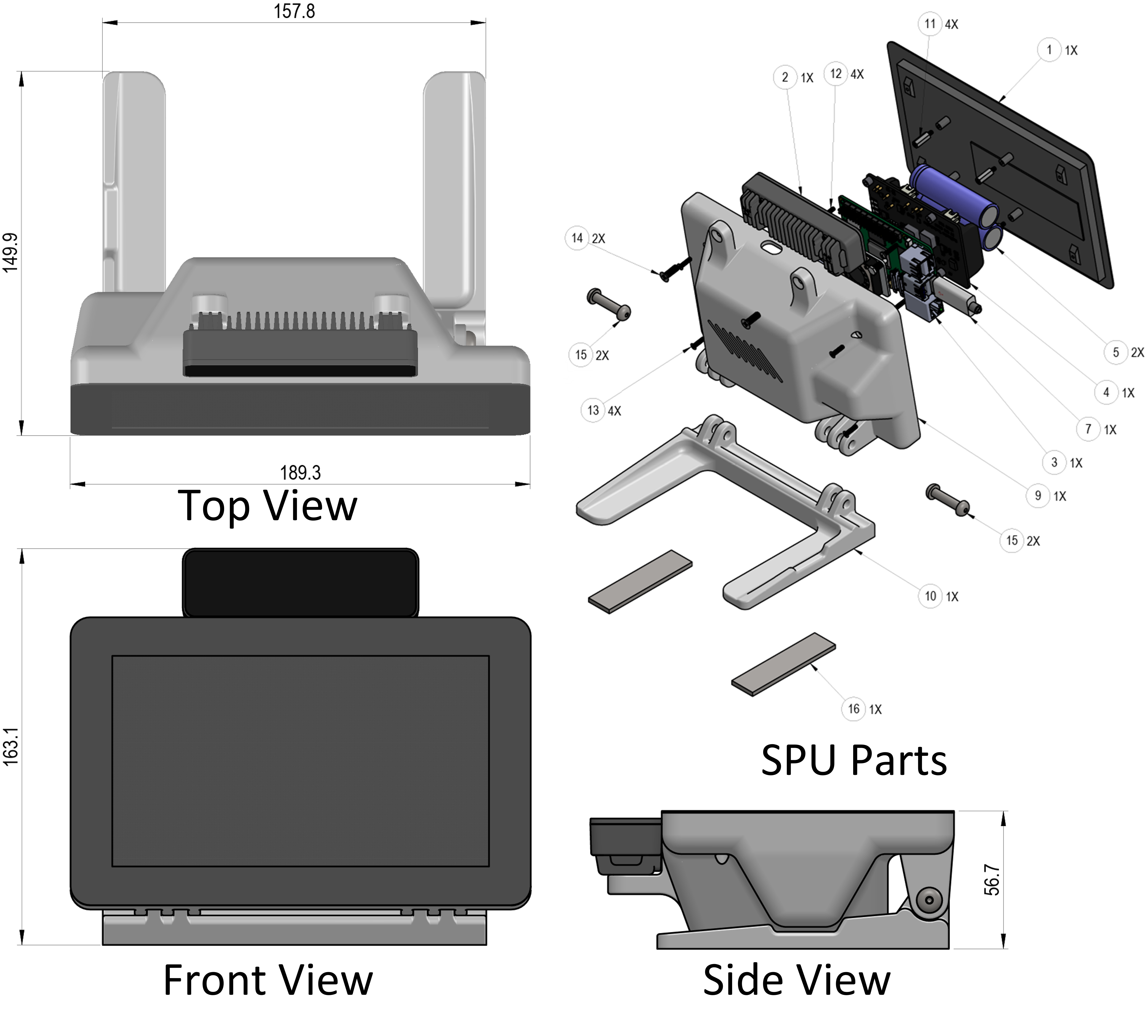}
    \caption{Mechanical drawings of the SPU enclosure showing key dimensions, mounting features, and component placement. All dimensions are given in millimeters. Refer to Table~\ref{tab:SPU_BOM} for the designated parts corresponding to the identifiers shown in the upper-right drawing.}

    \label{fig:SPU_MECH_DRAW}
\end{figure}

A custom enclosure was designed and fabricated to securely house all SPU electronic components while maintaining a compact and functional form factor. The enclosure was manufactured using the University of Louisville’s 3D-printing facilities. Its design satisfies several critical requirements: (i) it exposes only the USB-C charging port and the power button, ensuring that all unused Raspberry~Pi~5 ports remain physically inaccessible for enhanced hardware security; (ii) it provides an internal mounting structure that enables straightforward assembly of the SBC, camera, display, and battery system; and (iii) it incorporates a mechanical rotation-limiting feature for the touch display hinge, preventing the SPU from tipping forward when placed on a desk. The final assembled enclosure has overall external dimensions of approximately $19\times16.3\times5.7$\,cm, making it compact enough for unobtrusive placement on a student’s desk. Detailed mechanical drawings—including all dimensions and subcomponent placements—are provided in Figure~\ref{fig:SPU_MECH_DRAW}. Table~\ref{tab:SPU_BOM} lists all components referenced in the mechanical design.

\begin{figure}[h!]
    \centering
    \includegraphics[width=0.5\textwidth]{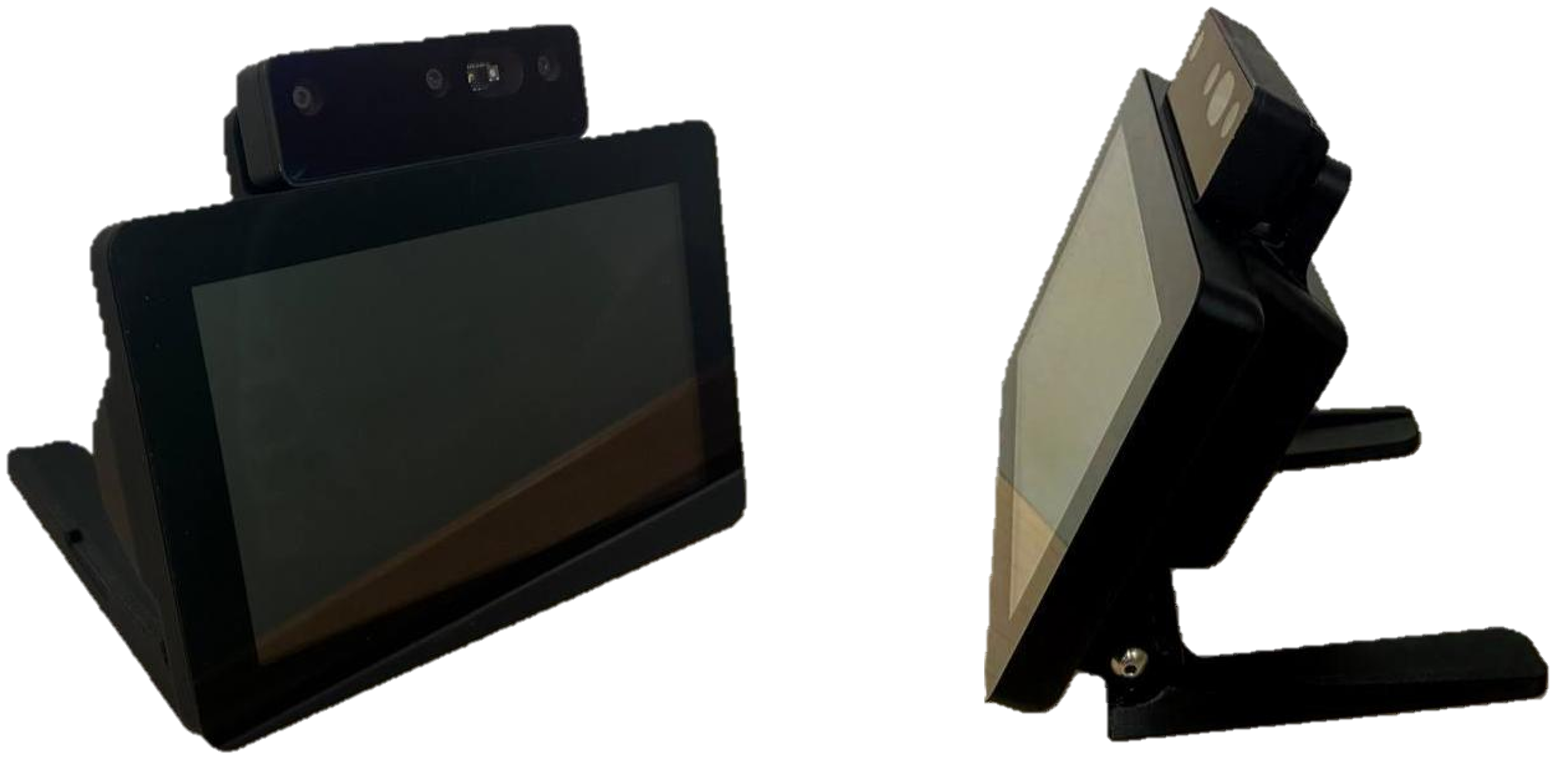}
    \caption{Fabricated SPU unit showing the assembled enclosure, integrated touch display, camera module, and internal power system. The image illustrates the final physical form factor and layout of the complete sensing node as deployed in classroom environments.}

    \label{fig:SPU_FAB_UNIT}
\end{figure}

\begin{table}[h!]
\centering
\renewcommand{\arraystretch}{1.25}

\resizebox{\textwidth}{!}{
\begin{tabular}{|p{1.0cm}|l|c|p{2cm}|}
\hline
\textbf{Item No.} & \textbf{Description} & \textbf{Qty} & \textbf{Unit Price (\$)} \\
\hline
1  & Raspberry Pi Touch Display~2 & 1 & 60 \\ \hline
2  & OAK-D Lite Camera Module & 1 & 149 \\ \hline
3  & Raspberry Pi~5 SBC (2.4~GHz, Quad-Core, 4~GB RAM) & 1 & 64.26 \\ \hline
4  & Geekworm X1200 UPS HAT (5~V / 5~A) & 1 & 43 \\ \hline
5  & 18650 Lithium-Ion Battery Pack (Set of 2) & 1 & 13 \\ \hline
6  & Raspberry Pi~5 Active Cooler & 1 & 5 \\ \hline
7  & Right-Angle USB-C Power Cable & 1 & 5 \\ \hline
9  & SPU Screen Enclosure (3D Printed) & 1 & 7 \\ \hline
10 & SPU Screen Bracket (3D Printed) & 1 & 3 \\ \hline
11 & Aluminum Male–Female Threaded Hex Standoff, M2.5  & 4 & 0.16 \\ \hline
12 & Stainless Steel Phillips Pan Head Screw, M2.5, 12mm & 4 & 0.05 \\ \hline
13 & Stainless Steel Phillips Flat Head Screw, M2.5, 10mm & 4 & 0.05 \\ \hline
14 & Stainless Steel Phillips Flat Head Screw, M4 & 2 & 0.09 \\ \hline
15 & Multipurpose 304 Stainless Steel Flat Bar & 2 & 0.83 \\ \hline
16 & M5 Binding Barrel and Chicago Screw (18-8 Stainless Steel) & 2 & 0.32 \\ \hline
17 & Class~10 MicroSD Card (64~GB) & 1 & 10.25 \\ \hline
\end{tabular}
}

\caption{Bill of Materials (BOM) for the SPU}
\label{tab:SPU_BOM}
\end{table}
A fully assembled SPU prototype was fabricated using the enclosure, electronic components, and mechanical integration described in the previous subsections. The completed unit demonstrates the final physical form factor, port layout, display mounting mechanism, and overall ergonomics of the device. Figure~\ref{fig:SPU_FAB_UNIT} illustrates the finished SPU assembly as deployed in classroom environments.

Having established the complete hardware architecture of the SPU and detailed the design and fabrication of its enclosure, we now turn to the software stack that enables the device to operate as an intelligent sensing node within the BSN. While the hardware provides the necessary physical platform for computation, sensing, and power management, it is the software that governs session control, secure communication, on-device inference, and overall system reliability. The next section, therefore, outlines the SPU software architecture and the key components required to meet the functional, security, and real-time processing requirements defined in the previous chapter.

\begin{sidewaysfigure}
    \centering
    \includegraphics[width=0.95\textheight]{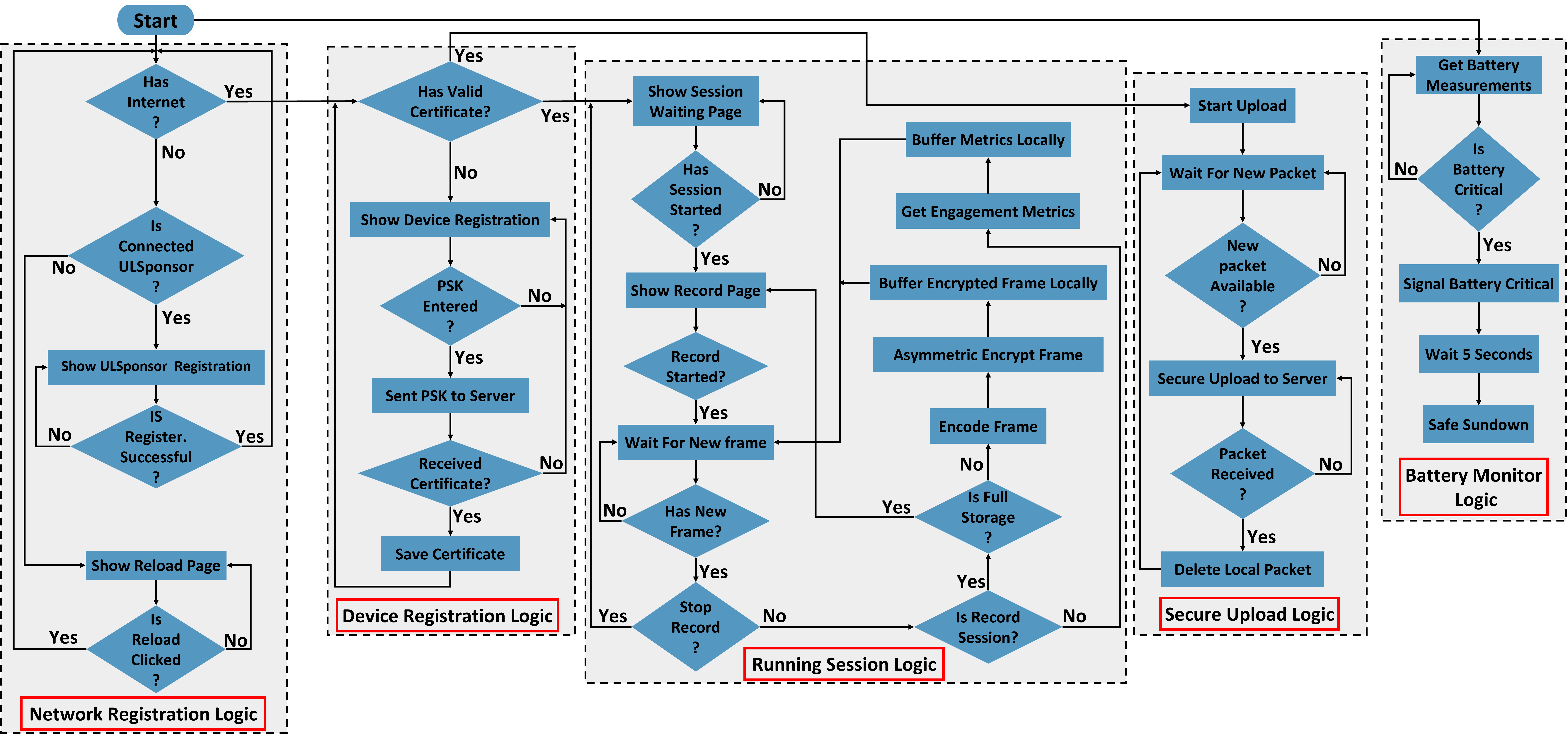}
    \caption{Software architecture flowchart of the SPU, illustrating the five primary subsystems: network registration, device registration, session management, upload logic, and battery monitoring. This diagram summarizes the complete operational workflow executed by the SPU from boot to shutdown.}
    \label{fig:SPU_SW_FLOWCHART}
\end{sidewaysfigure}

\section{SPU Software}

The SPU software architecture is composed of five tightly integrated subsystems that together enable secure, autonomous, and real-time operation in classroom environments. As illustrated in Figure~\ref{fig:SPU_SW_FLOWCHART}, these subsystems manage Wi-Fi onboarding, cryptographic device registration, session execution and consent handling, secure data upload, and continuous battery health monitoring. Although not shown explicitly in the flowchart, the SPU GUI also includes a persistent status bar that displays the system clock, battery level, Wi-Fi state, server connectivity, certificate validity, and quick-access controls for settings and power management. 

Each of the five subsystems is described in detail in the following subsections. Before doing so, we briefly outline the underlying software technologies that support their implementation and ensure consistent, reliable operation across all SPU devices.

\subsection{Software Technology Stack}

The SPU software is implemented using a modern and lightweight technology stack designed for robustness, portability, and efficient execution on embedded hardware. Python\cite{Python} serves as the primary development language due to its readability and rich ecosystem of CV and ML libraries. The user interface is built using PySide6 \cite{QtForPython6Docs}, providing a responsive, touch-friendly GUI suitable for classroom deployment.

Camera communication and on-device inference are handled through the DepthAI framework\cite{DepthAI_Docs}, which interfaces seamlessly with the OAK-D Lite to access RGB streams, stereo depth, and neural-network outputs. Secure uploads to the server’s S3-compatible object storage are performed using the \texttt{boto3} library\cite{Boto3Docs}, ensuring encrypted and reliable data transfer.

To support maintainable and reproducible development, the project employs a modern continuous integration and continuous delivery (CI/CD) toolchain, including Git with GitHub Actions for automated builds, Docker for environment isolation, \texttt{pytest}\cite{pytest_docs} for unit and integration testing, and \texttt{ruff}\cite{RuffDocs} for linting and code-quality enforcement. Python environments are managed using miniforge \cite{Miniforge}, which enables consistent dependency resolution across development machines and ARM-based deployment targets such as the Raspberry~Pi~5. Together, this technology stack provides a stable, scalable, and developer-friendly foundation for implementing, validating, and deploying the SPU software.

\subsection{Network Registration Logic}

The SPU is designed to operate primarily on the University of Louisville’s \texttt{ULSponsor} wireless network. Relying on this campus-wide onboarding network provides several advantages: (i) all SPU traffic is automatically protected by the university’s Virtual Private Network and firewall infrastructure; (ii) the SPU can be deployed in any classroom on campus without additional configuration; and (iii) no private access points or custom network setups are required. Although the SPU is technically capable of connecting to any Wi-Fi network, this capability is reserved for administrative debugging and testing and is not exposed through the SPU GUI. Manual network changes are performed only by an authorized system administrator.

Upon startup, the SPU begins by checking for Internet connectivity. If connectivity is unavailable, it determines whether it is currently associated with the \texttt{ULSponsor} network. When connected to this onboarding SSID, the SPU displays a registration interface that allows an authorized user to authenticate the device and request Wi-Fi access for its MAC address. This process must be performed only once per SPU---as long as the device retains the same MAC address, it remains authorized on the university network. To support this workflow, the system administrator must first create a guest account, which is then used to onboard any new SPU device.

If registration is successful, the SPU reloads its network configuration and repeats the connectivity check. If registration fails, the device continues prompting for onboarding until Internet access is available. This subsystem ensures that each SPU can securely join the campus network without requiring per-classroom setup or manual configuration, enabling fully autonomous operation across all deployment locations.

\begin{figure}[t]
    \centering
    \includegraphics[width=\textwidth]{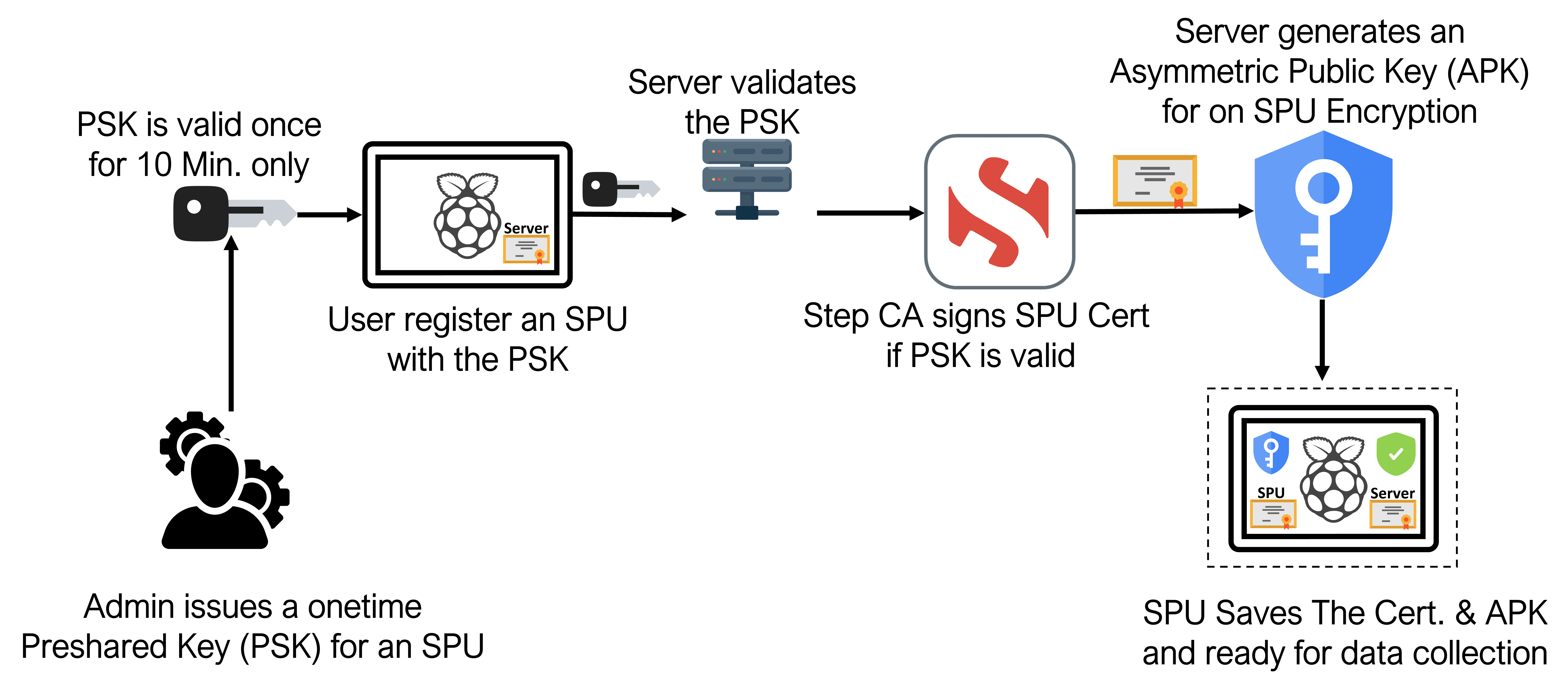}
    \caption{Device registration flow for the SPU. The diagram illustrates the certificate-based onboarding process, including network verification, PSK-based authorization, CSR generation, certificate issuance via STEP~CA, and the return of the server-generated public key used for secure local data encryption.}
    \label{fig:SPU_SW_DEV_REG}
\end{figure}

\subsection{Device Registration Logic}

Before an SPU can participate in classroom deployments, it must be cryptographically authenticated and registered with the BSN server. As illustrated in Figure~\ref{fig:SPU_SW_DEV_REG}, each SPU is provisioned at manufacture time with the server’s public certificate, ensuring that all subsequent communication occurs only with the legitimate BSN server and preventing man-in-the-middle and spoofing attacks. Upon startup, the SPU checks whether a valid device certificate is already present. If so, the device bypasses registration and proceeds directly to the session-waiting state. If no valid certificate exists, the SPU displays the device-registration interface, which prompts the user to enter a time-limited pre-shared key (PSK). This PSK is generated by the system administrator for a specific classroom, is valid for only ten minutes, may be used exactly once, and ensures that each SPU is bound to a single classroom at any given time.

Once the PSK is entered, the SPU locally generates a private key and a certificate signing request (CSR). The CSR, along with the PSK and the device identifier, is sent to the BSN server. The server verifies the PSK and, if valid, signs the CSR using the on-premises STEP CA, thereby issuing a device certificate. In addition, the server generates a dedicated asymmetric key pair for secure data protection: the private key is stored exclusively on the server and never leaves it, while the corresponding public key is returned to the SPU. This public key is later used by the SPU to encrypt buffered engagement data stored locally before upload, ensuring that even if an SPU is stolen or compromised, the data cannot be decrypted—only the BSN server, which holds the matching private key, can recover it.

After receiving the signed certificate and the server-side public key, the SPU stores both credentials securely and transitions into normal operational mode. If PSK verification fails, the device remains on the registration screen until valid credentials are provided. This certificate-based onboarding procedure needs to be performed only once—typically by
an administrator or a designated technical staff member—before the SPU is deployed in a classroom. Once completed, the SPU is fully authenticated, cryptographically trusted, and able to store sensitive data securely, enabling it to integrate seamlessly into the
classroom sensing workflow.

\begin{figure}[t]
    \centering
    \includegraphics[width=0.95\textwidth]{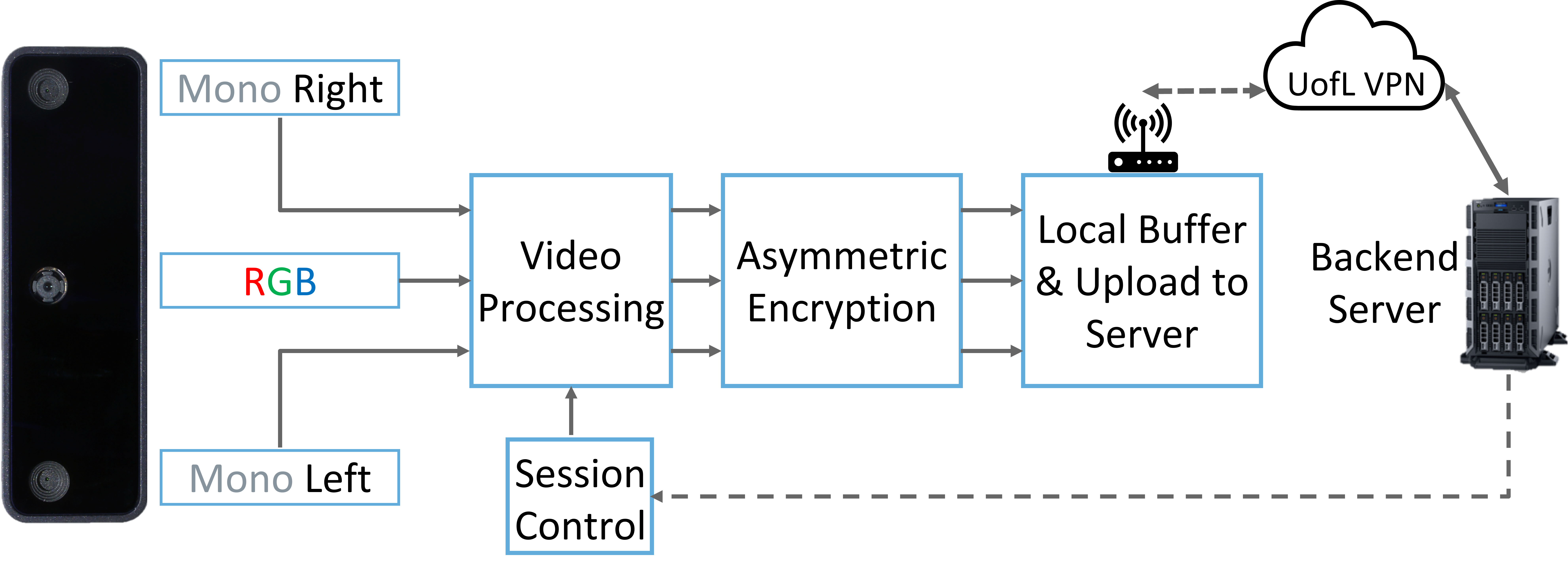}
    \caption{Functional block diagram of the SPU recording-session logic, showing the sequential workflow for session control (start/stop based on server signals or user interaction), video capture and encoding, hybrid encryption, local buffering and uploading, and session termination.}
    \label{fig:SPU_SW_SESS_REC}
\end{figure}

\subsection{Running Session Logic}

Once device registration is completed, the SPU transitions into a waiting state in which it periodically polls the BSN server for the activation of a classroom session. The system supports two distinct session types: a \emph{record session} used during the development phase to collect raw video for dataset construction, and an \emph{analysis session} used during deployment for real-time engagement inference. When the server signals the start of a session, the SPU presents a consent interface to the student. In compliance with IRB requirements, the camera pipeline remains fully disabled until the student explicitly grants consent. Only after consent is obtained does the SPU enter the active sensing workflow appropriate for the selected session type.

For a \textbf{recording session}, the SPU captures video frames from the OAK-D~Lite at 25~fps, using a 720p resolution stream from the RGB camera and a 480p stream from the mono stereo pair. Each frame is first encoded on the camera’s hardware accelerator using the H.264/H.265 codec and then encrypted before being written to the local buffer.

To minimize computational overhead, the SPU employs a hybrid encryption scheme in which each encoded frame is encrypted using a randomly generated SHA-256–based symmetric key \cite{FIPS1804}, while the symmetric key itself is encrypted using the server’s asymmetric public key. The encrypted frame, encrypted symmetric key, and associated metadata are then written to the local buffer, where they remain only temporarily: as soon as network conditions allow, the SPU uploads each buffered frame to the server’s S3-compatible object storage \cite{AmazonS3} and deletes the local copy immediately upon confirmed receipt. Each record includes session identifiers, timestamps, camera intrinsics, frame dimensions, and device identifiers, following the structure:

\begin{verbatim}
{
    "session_id": ...,
    "video_id": ...,
    "frame_counter": ...,
    "metadata": ...,
    "encrypted_key": ...,
    "nonce": ...,
    "tag": ...,
    "device_id": ...,
    "timestamp": ...,
    "device_timestamp": ...,
    "timestamp_offset": ...,
    "width": ...,
    "height": ...,
    "type": ...
}
\end{verbatim}

For an \textbf{analysis session}, the SPU processes video frames on the edge to compute engagement metrics in real time. Frames are grouped into 10-second windows, and for each window, the SPU performs face detection, gaze estimation, and behavioral feature extraction to derive an engagement score. Both the encrypted frames and the resulting engagement metrics are buffered in local storage. The SPU continues this process until the session ends or the student manually stops recording. All buffered data is subsequently transmitted to the server by the upload subsystem, which handles secure delivery and confirmation-based deletion.

The overall workflow executed during a recording session is summarized in the functional block diagram shown in Figure~\ref{fig:SPU_SW_SESS_REC}, which illustrates the control flow governing session activation, consent handling, frame acquisition, encryption, buffering, and termination conditions.

\subsection{Secure Upload Logic}

\begin{figure}[t]
    \centering
    \includegraphics[width=0.95\textwidth]{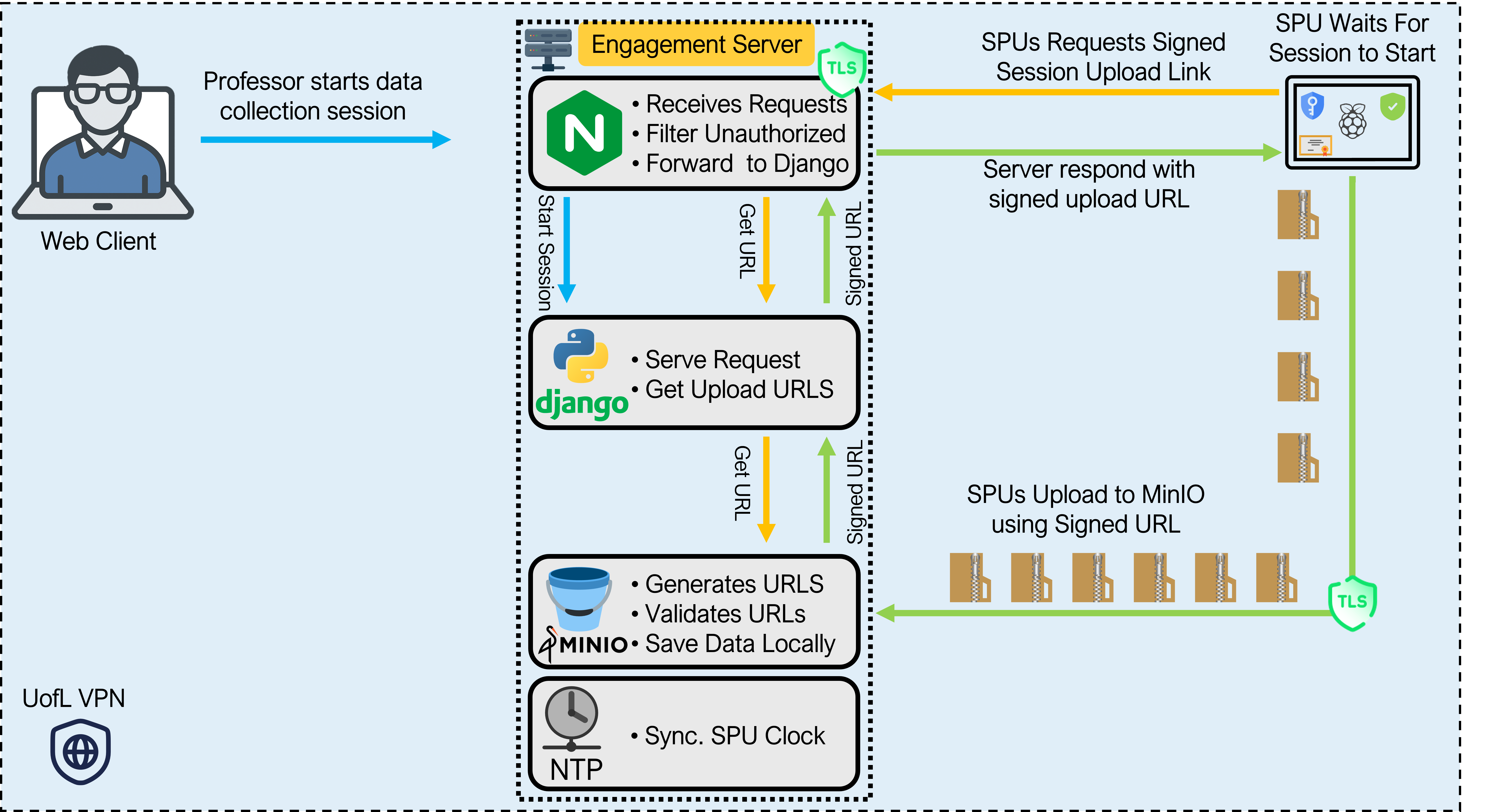}
    \caption{Functional block diagram illustrating the interaction between the SPU and the BSN backend server. The BSN server stack consists of an NGINX reverse proxy, a Django application server, an on-premises MinIO object-storage service, and an NTP time synchronization service. The diagram shows how the SPU requests time-limited, upload-only pre-signed URLs and transmits encrypted data to the server over mutually authenticated TLS~1.3 connections.}
    \label{fig:SPU_SERVER_ARHC}
\end{figure}

The upload subsystem operates continuously in parallel with all other SPU processes once the device possesses a valid certificate. This ensures that any previously unuploaded packets—including those generated during temporary network outages—are eventually delivered reliably. The overall workflow is illustrated in Figure~\ref{fig:SPU_SERVER_ARHC}, which summarizes the interaction between the SPU, the BSN server, and the on-premises MinIO storage service.

At regular intervals, the SPU requests from the BSN server a time-limited, upload-only pre-signed link to the MinIO object-storage backend \cite{MinIO}. The server verifies the SPU’s certificate and, if valid, issues a pre-signed URL that remains active for 30 minutes. The SPU uses this URL to upload any buffered packets during that window; once the link expires, the SPU automatically requests a new one to maintain continuous authorization without exposing long-lived credentials.

All uploads are performed over a mutually authenticated TLS~1.3 \cite{RFC8446} connection. Upon successful receipt of each packet (or batch), the server sends an acknowledgment, after which the SPU deletes the local copy to prevent redundant retransmissions and reduce storage usage. If a packet fails to upload, it remains in the buffer until connectivity is restored or a new upload window becomes available.

To avoid overloading the server with large numbers of individual upload requests—especially during recording sessions where each video frame is encrypted independently—the SPU aggregates encrypted frames into ZIP archives of fifty frames each. Every ZIP bundle contains encrypted frames, associated metadata, and the corresponding encrypted symmetric keys. This batching strategy significantly improves throughput, reduces request overhead, and preserves confidentiality and integrity guarantees throughout the upload process.

\subsection{Battery Monitor Logic}

The SPU continuously monitors its battery voltage and charge percentage through the UPS module’s fuel-gauge interface. When the battery voltage falls below 3 V, the SPU issues a ``battery-critical'' signal to the session-control subsystem, allowing several seconds for pending uploads to complete and local buffers to flush before initiating a controlled shutdown. This mechanism prevents abrupt power loss, protects the battery from over-discharge, mitigates the risk of filesystem corruption, and ensures that no partially written data remains in an inconsistent state. The battery-monitor logic operates independently of both the recording and upload loops, providing continuous, power-aware supervision throughout the entire session.

\subsection{SPU User Interface Workflow}
\begin{sidewaysfigure}
    \centering
    \includegraphics[width=0.95\textheight]{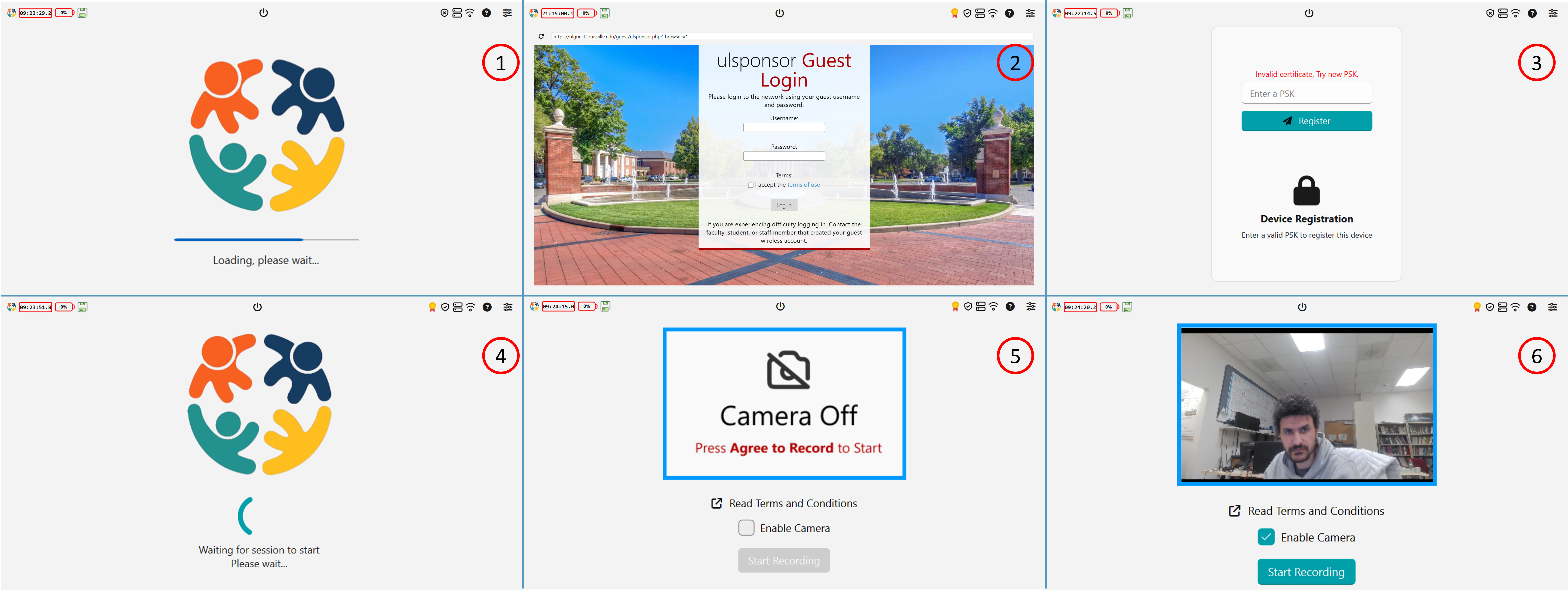}
    \caption{User-interface flow of the SPU software, shown as six sequential screens: (1) splash/loading screen; (2) ULSponsor Wi-Fi registration page shown when the device has no Internet access; (3) device-registration screen for PSK-based certificate provisioning; (4) session-waiting interface displayed after successful onboarding; (5) recording interface shown at session start with the camera disabled pending student consent; and (6) active recording interface displayed once consent is granted.}
    \label{fig:SPU_SW_PAGES}
\end{sidewaysfigure}

Figure~\ref{fig:SPU_SW_PAGES} illustrates the complete user-interface flow of the SPU application, shown as a sequence of six screens arranged in a 2$\times$3 layout. Screen~1 displays the loading and splash interface that appears immediately after the
SPU boots. If the device lacks Internet connectivity and is associated with the``ULSponsor'' onboarding network, Screen~2 is presented, prompting the user to enter valid UL credentials to obtain Wi-Fi access. Once network connectivity is established, the SPU checks whether a valid device certificate is installed. If no certificate is found, Screen~3---the device-registration interface---is shown, where the user must enter a time-limited pre-shared key (PSK) to authorize provisioning.

Upon successful certificate issuance, the SPU transitions to Screen~4, the session-waiting state, where it polls the BSN server for an active classroom session. When a session begins, the SPU displays Screen~5, the recording interface with the camera disabled until student consent is explicitly granted. After the student provides consent, the camera pipeline is activated and the SPU enters Screen~6, the full recording interface used for either dataset collection or engagement-analysis sessions. This sequence ensures secure, ethical, and controlled activation of the SPU in accordance with IRB and system requirements.

\subsection{SPU Operating System Security}

To ensure that each SPU operates as a secure, tamper-resistant sensing node, the device runs on a hardened Linux-based operating system that has been explicitly configured to minimize its attack surface. All nonessential services and interfaces are disabled, including SSH access, physical TTY terminals, and the standard desktop environment. As a result, even if an attacker connects an external keyboard or mouse to the SPU, the device cannot be interacted with or placed into a standard login shell.

Instead of relying on conventional user login workflows, the SPU launches directly into its application runtime under a restricted system user. The graphical interface is served through a minimal X11 environment without any window manager, shell access, or desktop session, thereby preventing privilege escalation through GUI-based interactions. The filesystem is locked down to prevent modification of system binaries, and all software updates are delivered manually.

Together, these measures create a tightly controlled execution environment in which the SPU behaves exclusively as an appliance: it performs its sensing, inference, and upload tasks autonomously, while offering no general-purpose computing capabilities to a physical attacker. This hardened OS configuration is essential for maintaining data privacy, preventing unauthorized access, and ensuring the integrity of the BSN deployment across classroom environments.

\section{Server Software}

Django was selected as the primary backend framework for the BSN because it provides a mature, secure, and highly extensible web-application foundation while allowing the entire backend to be implemented in Python. This minimizes the number of programming languages required across the BSN ecosystem, simplifies development and maintenance, and enables seamless integration with the Python-based SPU software stack. Figure~\ref{fig:SPU_SERVER_ARHC} presents a high-level block diagram of the BSN server architecture, illustrating how instructors interact with the system through a web client and how SPUs communicate with the backend during data-upload sessions.

The backend of the BSN system is implemented as a Django-based web server\cite{Django} with a MySQL\cite{MySQL} relational database and a modular, microservice-oriented architecture. Each functional domain of the system---user authentication, classroom configuration, device provisioning, session coordination, and secure data upload---is encapsulated in a dedicated Django app. The server is deployed using Gunicorn as the WSGI application server \cite{Gunicorn} and NGINX\cite{NGINX} as a reverse proxy, providing scalability, security, and production-grade performance.

The backend runs on an Ubuntu~20.04\cite{Ubuntu2004} server installation and integrates with two critical on-premises services: (i) a STEP~CA\cite{StepCA} instance responsible for issuing and revoking SPU certificates, and (ii) a MinIO S3-compatible object-storage service\cite{MinIO} that manages encrypted data ingestion. All server functionality is exposed through a versioned REST API \cite{Fielding2000} under the \texttt{/api/v1/} namespace. The following subsections describe each major backend component and its role within the overall system architecture.

\subsection{Accounts App}

The \textit{accounts} app provides user authentication, authorization, and identity management for instructors, researchers, and administrators. Built on top of \texttt{dj-rest-auth}, it exposes endpoints for login, logout, password reset, email verification, and user profile retrieval. Role-Based Access Control (RBAC) ensures that only authorized users can initiate sessions, configure classrooms, or access sensitive data. The primary API endpoints provided by this app are listed below:
\begin{verbatim}
/api/v1/dj-rest-auth/
/api/v1/dj-rest-auth/user/
\end{verbatim}

\subsection{Classroom Manager App}

The \textit{classroom\_manager} app maintains metadata describing classrooms, their assigned SPUs, and operational configurations. Administrators may retrieve classroom records, update settings, and bind SPUs to specific physical environments. This mapping is essential for IRB compliance and for contextualizing engagement data. The primary API endpoints provided by this app are listed below:

\begin{verbatim}
/api/v1/classroom/
/api/v1/classroom/<pk>/
\end{verbatim}

\subsection{Device Authentication App}

The \textit{device\_auth} app implements the certificate-based onboarding workflow used to authenticate each SPU before it enters operational mode. Its responsibilities include: (i) validating time-limited pre-shared keys (PSKs); (ii) receiving Certificate Signing Requests (CSRs); (iii) invoking STEP~CA to issue signed device certificates; and (iv) managing health-check telemetry. The primary API endpoints provided by this app are listed below:

\begin{verbatim}
/api/v1/device/provision/
/api/v1/device/health/
/api/v1/psk/
/api/v1/psk/<pk>/regenerate/
\end{verbatim}

This subsystem ensures that every SPU is cryptographically trusted and securely provisioned prior to participating in the sensing workflow.

\subsection{Session Manager App}

The \textit{session\_manager} app orchestrates the lifecycle of classroom sessions, including start and stop events, session-type configuration (recording or analysis), and real-time polling by SPUs. When a session begins, the server provides the SPU with the necessary session metadata; when it ends, the server signals termination so that the device can finalize encoding, flush buffers, and stop sensing. The app also ensures that only one active session may exist for a given classroom at any time, preventing conflicting operations. Furthermore, sessions are automatically deactivated once their predefined duration has elapsed, ensuring consistent system behavior even if the instructor forgets to stop the session manually. The primary API endpoints provided by this app are listed below:

\begin{verbatim}
/api/v1/session/start/
/api/v1/session/stop/
/api/v1/spu/session/current/
\end{verbatim}

\subsection{Storage App}

The \textit{storage} app handles integration with the on-premises MinIO server used for encrypted data ingestion. It generates time-limited, upload-only pre-signed URLs that SPUs use to transmit encrypted bundles of data. Before a URL is issued, the server verifies the device's certificate to ensure proper authentication. The primary API endpoints provided by this app are listed below:

\begin{verbatim}
/api/v1/spu/storage/presigned-directory/
\end{verbatim}

This approach enforces a zero-trust model: SPUs can only upload, never read or delete objects, and their upload permissions expire automatically (typically in 30 minutes).

\subsection{API Security and Transport Layer Protection}

All server endpoints are protected using TLS~1.3 to ensure the confidentiality and integrity of all communications between SPUs, user clients, and the backend services. In addition to transport-layer encryption, virtually all SPU-facing APIs require the device to present a valid, server-signed client certificate as part of a mutual TLS (mTLS) handshake. This certificate-based authentication prevents unauthorized or spoofed devices from interacting with the system, blocks man-in-the-middle attacks, and enforces a strict zero-trust communication model. As a result, only properly provisioned SPUs are able to register, join sessions, upload encrypted data, or retrieve operational configuration from the BSN server.

\subsection{Clock Synchronization}

To ensure temporal consistency across all SPUs during classroom operation, the backend server also functions as the authoritative time source for the BSN. The server hosts a local NTP service\cite{RFC5905} to maintain a stable and accurate system clock and exposes this reference to all connected SPUs. During normal operation, each SPU periodically synchronizes its local clock with the server’s timestamp, achieving a typical alignment accuracy within 200--500~ms. This level of precision is sufficient for aligning engagement metrics, session boundaries, and video-frame timestamps across devices, and is essential for both dataset construction and real-time classroom analytics. By centralizing time synchronization through an on-premises NTP server cite{RFC5905}, the BSN maintains coherent temporal semantics even in the presence of network jitter or device-specific clock drift.

\subsection{Calibration}
\begin{figure}[p] 
    \centering
    \includegraphics[width=\textwidth]{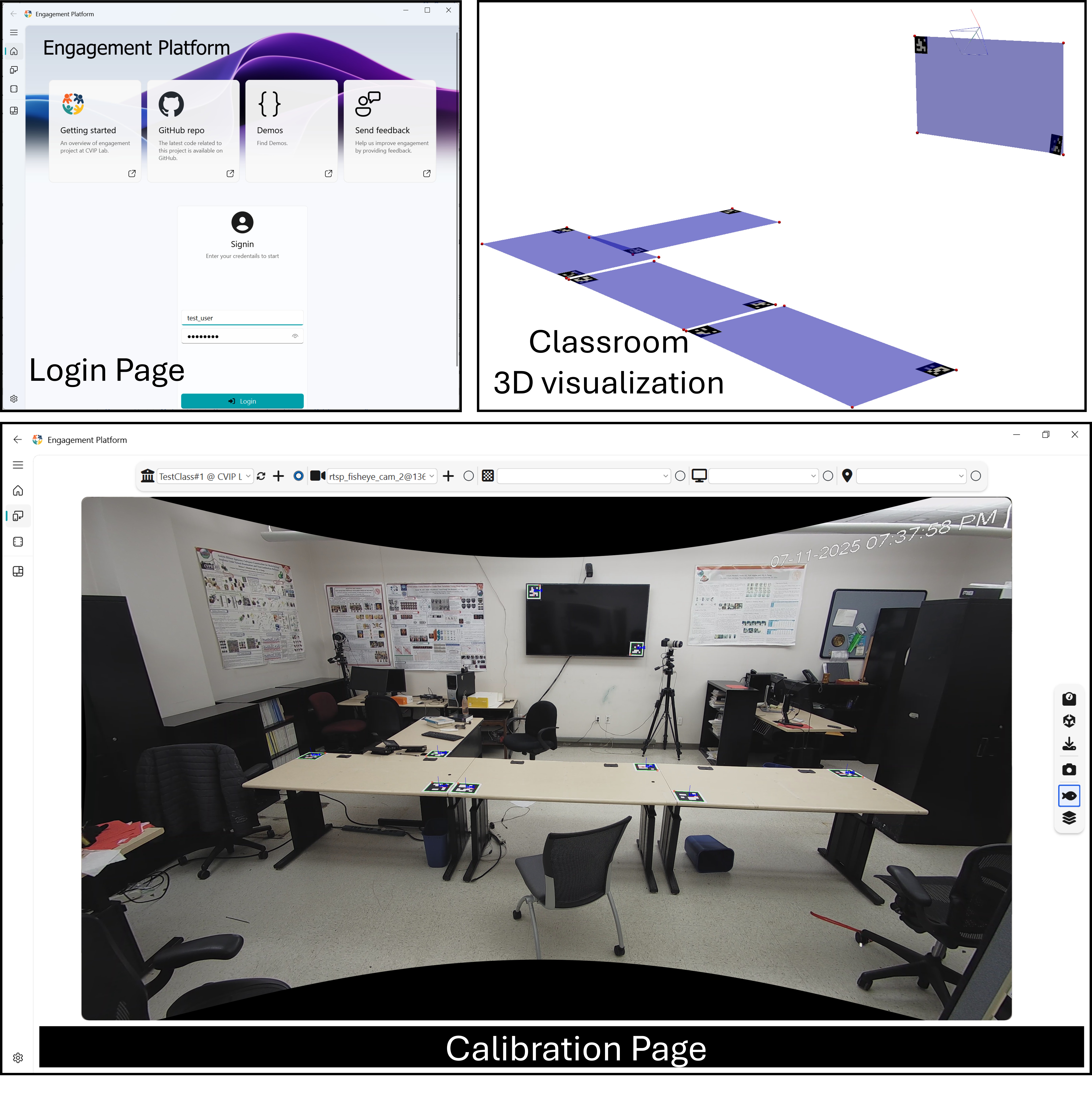}
    \caption{Three windows of the developed front-end GUI for classroom calibration. The top-left window is the login page. The bottom window is the calibration page, which supports adding new classrooms and cameras, performing calibration, and visualizing the results. The image in the center of the bottom window shows a live camera stream that has been rectified to remove distortion. The top-right window presents a 3D visualization of a calibrated classroom at the CVIP Lab. Although the classroom contains a single actual screen, additional dummy surfaces (e.g., desk surfaces) were added as objects of interest to test the calibration algorithm.}
    \label{fig:CALIB_GUI}
\end{figure}
One advantage of using Oak camera is it comes with its cameras are calibrated both for stereo and RGB. This eliminates the need to calibrate each SPU manually, which is a tedious process. For the FCUs we developed a GUI that can be used to help make calibration of these cameras easy. This GUI  supports adding new classrooms and cameras, performing calibration, and visualizing the results. Moreover this GUI also supports calibrating objects of interest, such as a blackboard or a projector screen. The calibration approach is simple: a fixed set of fiducial markers, e.g, ARUCO markers \cite{Garrido-Jurado2014}, placed in a classroom such that each pair of cameras sees at least one common marker. A pose graph is then created from the cameras and markers and optimized to get each camera pose relative to a common reference. A similar approach can also be used to mark the fixed objects of interest in the classroom, then a pose graph is created and optimized to obtain the pose of each object relative to a common reference in 3D. Each object of interest is modeled as a rectangular surface with two ARUCO markers placed at opposite corners, and it must be fully visible in at least one camera view. Our current system supports this feature; more specifically, we implemented the approach proposed by \cite{Garca-Ruiz2024} to calibrate our classroom. Figure~\ref{fig:CALIB_GUI} shows three windows from the front-end application used for classroom calibration and visualization of the results. To assess the performance of the calibration algorithm, we use the mean re-projection error, which quantifies the average discrepancy between the projected 3D points and their detected 2D locations in the camera images.

\chapter{BSN Evaluation}

This chapter presents a comprehensive evaluation of the proposed SPU and BSN architecture. The assessment serves two purposes: first, to compare the performance, security, and functional capabilities of the new SPU against the baseline system previously developed in the CVIP Lab and reported in \cite{Alkabbany2023}; and second, to demonstrate through a series of empirical tests that the proposed system meets the operational requirements defined in earlier chapters. The evaluation examines several critical dimensions of system performance, including battery-powered runtime (with a target of exceeding two hours of continuous operation), sustained network bandwidth for encrypted video-data transfer, clock synchronization accuracy between SPUs and the BSN server, and the SPU’s ability to execute moderate real-time CV/ML workloads such as face detection, gaze estimation, and emotion classification models. Together, these results validate that the proposed SPU achieves the intended design objectives and constitutes a substantial advancement over the earlier baseline platform.

\begin{table}[h!]
\centering
\renewcommand{\arraystretch}{1.25}
\begin{tabular}{|p{2cm}|p{5cm}|p{5cm}|}
\hline
\textbf{Feature} & \textbf{Baseline } \cite{Alkabbany2023} & \textbf{Proposed SPU} \\
\hline

SBC &
Raspberry Pi~4 &
Raspberry Pi~5 \\
\hline

CPU &
Quad-core \newline Cortex-A72 @ 1.5\,GHz &
Quad-core  \newline Cortex-A76 @ 2.4\,GHz \\
\hline

RAM &
2 up to 8\,GB LPDDR4  &
4up to 16\,GB LPDDR4X \\
\hline

Storage &
32\,GB microSD card &
64\,GB microSD card \\
\hline

Display &
Raspberry Pi Touch \newline Display~1 (800×480) &
Raspberry Pi Touch \newline Display~2 (720×1280) \\
\hline

Camera &
Official Pi Camera Module (no depth, no accelerator) &
DepthAI OAK-D Lite \newline 4K RGB, stereo depth \newline 4~TOPS AI accelerator \\
\hline

Enclosure &
Off-the-shelf enclosure; ports exposed;
no physical tamper protection &
Custom 3D-printed enclosure; sealed ports;
physical security\\
\hline

In Power &
Wired charger; no UPS &
7 Ah Battery \newline up to 2~hours up time; \newline safe shutdown support \\
\hline

Touch &
10-point capacitive touch &
5-point capacitive touch \\
\hline

AI Acc. &
None &
OAK-D Lite (4~TOPS) \\
\hline

Boot\newline Time &
50 Sec &
28 Sec \\
\hline

Wi-Fi &
Wi-Fi 5 &
Wi-Fi 6  higher throughput \& stability \\
\hline

Size &
14\,$\times$\,22\,$\times$\,4\,mm &
16.5\,$\times$\,19\,$\times$\,5.5\,mm \\
\hline

Cost &
\$200 &
\$383 \\
\hline

\end{tabular}
\caption{Comparison between the baseline \cite{Alkabbany2023} and the proposed BSN SPU hardware. The proposed system provides significant improvements in compute capability, wireless connectivity, camera quality, physical security, autonomy, and AI processing capacity with a moderate cost increase.}

\label{tab:baseline_vs_proposed}
\end{table}

\begin{table}[h!]
\centering
\renewcommand{\arraystretch}{1.25}
\begin{tabular}{|p{2.5cm}|p{5.5cm}|p{5.5cm}|}
\hline
\textbf{Feature} &
\textbf{Baseline} \cite{Alkabbany2023} &
\textbf{Proposed SPU} \\
\hline

Operating \newline System &
Pi OS with default services enabled &
Hardened Pi OS; disabled SSH, TTYs, desktop login; \\
\hline

GUI \newline Framework &
Tkinter running inside a full desktop environment &
PySide6 (Qt 6) in minimal X11 session; no desktop, no shell; kiosk-secure runtime \\
\hline

Startup \newline Behavior &
Requires auto-login to launch application &
Direct boot into SPU app; no login required; locked-down execution environment \\
\hline

Security Model &
TLS~1.3 only; device certificates (Manual); no encrypted storage &
Mutual TLS~1.3, device certificates (STEP~CA), encrypted local storage (hybrid AES+RSA) \\
\hline

Data \newline Handling &
Raw video stored unencrypted; manual transfer &
All buffered data AES-encrypted per frame; S3 upload-only URLs; server-side decryption only \\
\hline

Data \newline Upload &
No object storage &
Automated upload via MinIO; \newline confirmation-based deletion \\
\hline

Inference &
Yes &
Supports it, Not implemented \\
\hline

Clock Sync.&
No synchronized timestamps &
NTP-based server–SPU synchronization (200–500\,ms) \\
\hline

Session \newline Management &
Automatic start/stop  &
Server-driven sessions; dual modes (recording/analysis); automatic transitions \\
\hline

Network Reg. &
Manually connects to Wi-Fi; insecure configuration possible &
Automatic via ULSponsor \\
\hline

Device \newline Reg. &
Manual &
Automatic; PSK-based provisioning. \\
\hline

\end{tabular}
\caption{Comparison of software features between the baseline \cite{Alkabbany2023} and the proposed BSN SPU software. The proposed software stack provides major improvements in security and data privacy to support IRB-compliant classroom operation.}
\label{tab:SW_baseline_vs_proposed}
\end{table}

\section{Baseline Comparison}

To contextualize these results, it is necessary to evaluate the proposed BSN against the baseline SPU described in \cite{Alkabbany2023}, which represents the previous system revision developed in the CVIP Lab. The new SPU introduces major enhancements across both hardware and software. On the hardware side, the design upgrades four key subsystems: the camera (from a basic Pi camera to a 4K DepthAI module with on-board AI acceleration), the power architecture (from wired-only to battery-powered operation), the compute capability (from a Pi~4 to a Pi~5 with an auxiliary neural accelerator), and the enclosure (from off-the-shelf housings with exposed ports to a custom, physically secured enclosure). On the software side, significant advances include a comprehensive data-security and privacy framework, certificate-based authentication, hardened OS configuration, encrypted local storage, and a fully automated sensing and upload pipeline. These combined improvements enable a more secure, autonomous, and scalable BSN suitable for real-world classroom deployment, motivating a structured comparison against the earlier baseline.

\subsection{Experiment~1: Server Load and Network Throughput Evaluation}

This experiment evaluates whether the backend server and network infrastructure can sustain
the expected data traffic generated by multiple SPUs during a live classroom session. A test
classroom was created and populated with ten registered SPUs. Because a typical lecture
lasts approximately 60 minutes, a 70-minute test session was created and activated through
the admin portal to provide additional buffer for initialization and shutdown overhead.

\subsubsection*{Theoretical Bandwidth Requirement}

Each SPU generates three packets per second, with an average packet size of approximately
1.5~MB. Thus, the per-device data rate is

\[
3 \times 1.5~\text{MB/s} = 4.5~\text{MB/s},
\]

which corresponds to

\[
4.5~\text{MB/s} \times 8 = 36~\text{Mb/s}.
\]

For a deployment of ten SPUs operating simultaneously, the aggregate upload bandwidth
requirement is therefore

\[
10 \times 4.5~\text{MB/s} = 45~\text{MB/s} \approx 360~\text{Mb/s}.
\]

This value represents the theoretical worst-case throughput that the server and network infrastructure must sustain during a live classroom session under continuous data generation and upload.

On each SPU, a synthetic data generator was deployed to emulate real device output, matching the frame rate, packet size, and metadata structure of actual encrypted SPU payloads. Because the upload subsystem runs continuously, all SPUs immediately began
sending synthetic packets through the complete secure upload pipeline using time-limited pre-signed URLs.

During the 70-minute session, the server was unable to ingest packets in strict real time at the full theoretical rate; however, the SPUs successfully buffered all generated packets locally without data loss. After the session ended, the SPUs continued uploading their buffered packets, completing the process approximately one hour later. This demonstrates that the system can tolerate transient bandwidth saturation by leveraging local buffering, eventual consistency, and the periodic refresh of upload credentials.

To verify correctness, the number of packets generated and uploaded by each SPU was recorded. Local storage usage was measured before and after the session to confirm that all buffered packets were deleted after successful upload. Table~\ref{tab:experiment1_packet_summary} summarizes the results.

This experiment confirms that, even under heavy multi-device load, the BSN backend can safely and reliably ingest all data without loss by combining edge buffering with secure upload retries.

\begin{table}[h!]
\centering
\renewcommand{\arraystretch}{1.25}
\scriptsize
\begin{tabular}{|c|c|c|c|c|}
\hline
\textbf{SPU ID} &
\textbf{Packets Generated} &
\textbf{Packets Uploaded} &
\textbf{Packet Loss} &
\textbf{Storage Change (MB)} \\ 
\hline

SPU--01 & 4200 & 4200 & 0 & 0 \\ \hline
SPU--02 & 4200 & 4200 & 0 & 0 \\ \hline
SPU--03 & 4200 & 4200 & 0 & 0 \\ \hline
SPU--04 & 4200 & 4200 & 0 & 0 \\ \hline
SPU--05 & 4200 & 4200 & 0 & 0 \\ \hline
SPU--06 & 4200 & 4200 & 0 & 0 \\ \hline
SPU--07 & 4200 & 4200 & 0 & 0 \\ \hline
SPU--08 & 4200 & 4200 & 0 & 0 \\ \hline
SPU--09 & 4200 & 4200 & 0 & 0 \\ \hline
SPU--10 & 4200 & 4200 & 0 & 0 \\ \hline

\end{tabular}
\caption{Summary of server load and network-throughput experiment for ten SPUs. All devices successfully transmitted all generated packets with zero loss. Uploads were not real-time but completed within approximately one hour after the session ended, confirming correct buffering, retry logic, and cleanup of local temporary storage.}
\label{tab:experiment1_packet_summary}
\end{table}

\subsection{Experiment~2: Clock Synchronization Accuracy}

This experiment evaluates whether all SPUs maintain consistent time alignment with the BSN server, a requirement for synchronizing video frames, engagement metrics, and session boundaries. ten SPUs were powered on, and the clock offset between each device and the server was measured once every ten seconds for a duration of ten minutes via SSH. Table~\ref{tab:clock_sync_results} summarizes the results, reporting the number of samples per device, along with the average, minimum, and maximum of the measured offsets.

\begin{sidewaysfigure}
    \centering
    \includegraphics[width=0.95\textheight]{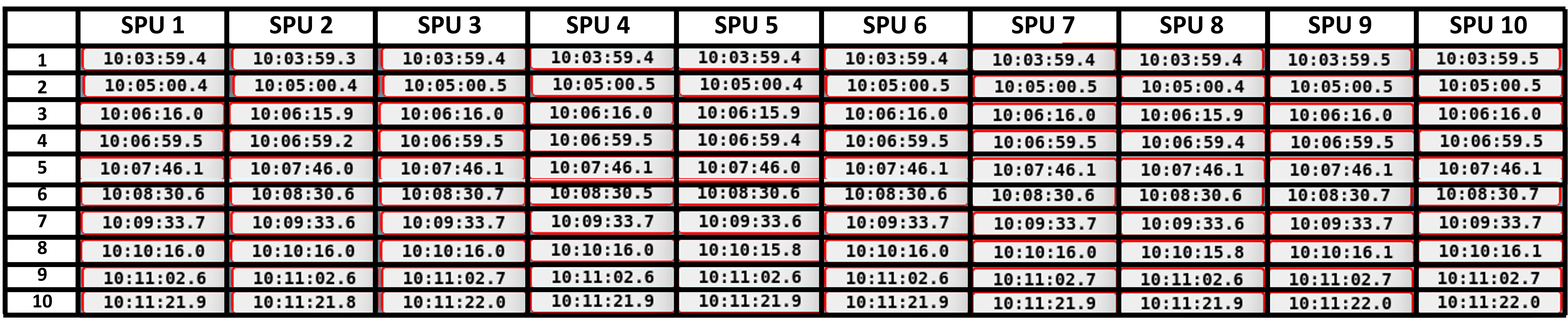}
    \caption{Visual synchronization check of ten SPUs during the clock-alignment experiment. The horizontal axis corresponds to SPU IDs, while the vertical axis represents time samples. Each cell shows the timestamp displayed by an SPU at a given moment. The near-identical timestamps across columns indicate that all SPUs remain tightly synchronized with the server throughout the ten-minute observation period.}
    \label{fig:SPU_SW_CLOCK}
\end{sidewaysfigure}

To further validate synchronization visually, the clock widget was remotely launched on ten SPUs, arranged in a grid layout on a laptop display, and recorded over a ten-minute period. Representative frames were extracted and inspected to verify that all SPUs' displayed times advanced in near-perfect unison. Figure~\ref{fig:SPU_SW_CLOCK} illustrates the captured grid view.

Although the timestamps displayed across SPUs are nearly identical, minor visual differences of up to approximately 100~ms can occasionally be observed between widgets in Figure~\ref{fig:SPU_SW_CLOCK}. These discrepancies do not reflect actual clock misalignment; rather, they arise from the rendering and transmission overhead introduced by the X11 forwarding pipeline used to display all clock widgets remotely on a single laptop screen. The underlying NTP-synchronized device clocks remain tightly aligned, as confirmed by the quantitative jitter measurements reported earlier.

Across all devices, the measured offsets fall within a narrow range (mostly within $\pm 3$~ms), with negligible jitter for the majority of SPUs. These results confirm that the NTP-based synchronization mechanism achieves the temporal precision required for classroom analytics and multi-device dataset alignment.

\begin{table}[h!]
\centering
\scriptsize
\renewcommand{\arraystretch}{1.25}
\begin{tabular}{|c|c|c|c|c|}
\hline
\textbf{Device ID} &
\textbf{Samples} &
\textbf{Avg Offset (ms)} &
\textbf{Min Offset (ms)} &
\textbf{Max Offset (ms)} \\
\hline

1  & 131 & -1.2096 & -2.4546 &  0.7432 \\ \hline
2  & 131 & -0.8002 & -0.8002 & -0.8002 \\ \hline
3  & 131 & -0.0242 & -0.0242 & -0.0242 \\ \hline
4  & 131 & -1.4537 & -1.4537 & -1.4537 \\ \hline
5  & 131 & -0.0757 & -0.0757 & -0.0757 \\ \hline
6  & 131 &  1.3071 &  1.3071 &  1.3071 \\ \hline
7  & 131 &  0.9944 & -0.6923 &  3.2164 \\ \hline
8  & 131 &  0.4537 &  0.4537 &  0.4537 \\ \hline
9  & 131 &  0.5046 &  0.5046 &  0.5046 \\ \hline
10 & 131 & -3.0781 & -3.0781 & -3.0781 \\ \hline
\end{tabular}
\caption{Clock synchronization offsets between each SPU and the BSN server measured over a ten-minute interval. The offsets show tight clustering around zero ms, confirming stable NTP-based synchronization across all devices.}
\label{tab:clock_sync_results}
\end{table}

\subsection{Experiment~3: Battery Endurance Under Continuous Operation}

To verify that the SPU can operate on battery power for the full duration of a typical lecture, battery-discharge tests were conducted under realistic sensing conditions. In the first trial, an SPU was fully discharged, then recharged completely while recording the total charging time. The full charging process required approximately two hours. A 70-minute session was then initiated, during which the device continuously recorded all three OAK-D Lite camera streams. Battery percentage before and after the session was logged.

A second trial repeated the test, this time starting from a 75\% charge level. Table~\ref{tab:battery_endurance} reports the session duration, initial and final battery percentages, and the effective battery consumption for both trials. These results confirm that the SPU can sustain more than two hours of continuous sensing, satisfying the system’s operational requirements.

\begin{table}[h!]
\centering
\caption{Summary of battery-endurance trials under continuous SPU operation.}
\renewcommand{\arraystretch}{1.25}

\resizebox{\textwidth}{!}{%
\begin{tabular}{|l|c|c|c|c|}
\hline
\textbf{Trial} &
\textbf{Session Duration (min)} &
\textbf{Initial Battery (\%)} &
\textbf{Final Battery (\%)} &
\textbf{Consumed (\%)} \\
\hline
Trial~1 (Full charge)   & 70 & 100 & 56 & 54 \\ \hline
Trial~2 (Partial charge) & 70 & 75  & 18 & 58 \\ \hline
\end{tabular}
}
\label{tab:battery_endurance}
\end{table}

\subsection{Experiment~4: On-Device AI Compute Performance}

This experiment evaluates whether the SPU’s integrated OAK-D~Lite accelerator can support the moderate real-time CV/ML workloads required for classroom engagement analysis. Three representative neural network pipelines were selected from the DepthAI model zoo, covering emotion recognition, gaze estimation, and fatigue detection. These tasks reflect the core sensing functions of the BSN. Each model was deployed on the SPU, and the achieved inference frame rate (FPS) was recorded. The results, summarized in Table~\ref{tab:ai_fps_table}, demonstrate that the OAK-D~Lite provides sufficient compute throughput to meet the latency and responsiveness requirements of the proposed system.

\subsubsection*{Emotion Recognition Pipeline}

The first evaluated model is the \textit{Emotion-Recognition-8-ENet} network \cite{savchenko2023facial}, based on the EfficientNet architecture \cite{tan2019efficientnet}. This model classifies faces into eight emotion categories: anger, contempt, disgust, fear, happiness, neutral, sadness, and surprise.  

On the OAK-D~Lite, the model operates as a two-stage DepthAI pipeline: (i) the YuNet face detector \cite{wu2023yunet} locates faces and produces crops; (ii) the ENet classifier evaluates each crop. The full pipeline achieves approximately 11~FPS using 64$\times$64 pixel face crops, enabling smooth multi-person classroom monitoring. A higher-resolution version of the model (256$\times$256 crops) was also tested and achieved approximately 5~FPS.

\subsubsection{Gaze Estimation Model}

The second evaluated model is the DepthAI \textit{Gaze Estimation} pipeline, which demonstrates multi-stage and multi-input inference on the OAK-D~Lite. The processing sequence consists of:

\begin{itemize}
    \item \textbf{Stage~1: Face Detection and Landmark Extraction} using YuNet \cite{wu2023yunet}, which provides facial keypoints for cropping the face and eyes.
    \item \textbf{Stage~2: Head Pose Estimation} using the OpenVINO head-pose model \cite{OpenVINO2023_head_pose_adas0001}.
    \item \textbf{Stage~3: Gaze Direction Estimation} using the ADAS gaze-estimation model \cite{OpenVINO2023_gaze_estimation}, combining cropped eye images with the estimated head-pose vector.
\end{itemize}

The complete pipeline achieves real-time performance at approximately 10~FPS on the OAK-D~Lite, making it suitable for estimating student visual attention during classroom instruction.

\subsubsection{Fatigue Detection Model}

The third evaluated model is the DepthAI \textit{Fatigue Detection} pipeline, which analyzes facial keypoints to estimate fatigue indicators such as eye closure and forward-leaning posture. After detecting and cropping the face from the RGB frame, the MediaPipe Face Landmarker model \cite{lugaresi2019mediapipe} produces a dense set of 3D facial keypoints.

These points are used to derive two key behavioral markers:

\begin{itemize}
    \item \textbf{Eye Closure:} Eye-aspect ratios are computed to detect prolonged eyelid closure.
    \item \textbf{Forward Leaning:} Geometric relationships among facial keypoints are analyzed to infer characteristic posture associated with fatigue.
\end{itemize}

The complete pipeline runs at approximately 10~FPS on the OAK-D~Lite, enabling reliable real-time detection during classroom sessions.

\begin{table}[h!]
\centering
\renewcommand{\arraystretch}{1.25}
\begin{tabular}{|l|l|p{4.0cm}|}
\hline
\textbf{Model / Pipeline} & \textbf{FPS} & \textbf{Notes} \\ \hline

Emotion Recognition (64$\times$64 crops) & 11 & Two-stage pipeline: YuNet + ENet classifier \\ \hline
Emotion Recognition (256$\times$256 crops) & 5 & Higher-resolution variant for improved accuracy \\ \hline
Gaze Estimation Pipeline & 10 & Three-stage pipeline: face detection, head pose, gaze estimator \\ \hline
Fatigue Detection Pipeline & 10 & MediaPipe Face Landmarker + posture/eye analysis \\ \hline

\end{tabular}
\caption{Inference performance of selected DepthAI pipelines running on the OAK-D~Lite accelerator. Results indicate that the accelerator provides sufficient throughput for real-time classroom engagement sensing.}
\label{tab:ai_fps_table}
\end{table}

\chapter{Conclusions}

This thesis presented the design, implementation, and evaluation of a novel Biometric Sensor Network (BSN) to enable real-time measurement of individual student engagement in STEM classroom environments. The system integrates robust hardware, secure software, computer-vision pipelines, wireless networking, and backend infrastructure into a unified platform capable of operating unobtrusively, ethically, and autonomously during live instructional sessions. 

The Student Processing Unit (SPU), developed as the foundational sensing node in the BSN, was engineered to meet five key system requirements: non-intrusive operation, non-invasive sensing, non-stigmatizing deployment, real-time processing, and fully automated operation. The redesigned SPU introduced several advancements over the previous baseline platform, including a significantly more powerful Raspberry~Pi~5 compute module, an OAK-D Lite camera providing 4K RGB imaging, stereo depth estimation, and a 4-TOPS AI accelerator, a battery-powered architecture that supports portable two-hour operation, and a custom 3D-printed enclosure offering enhanced physical security for classroom use. 

On the software side, the SPU incorporates a hardened operating system with disabled shells, restricted user access, encrypted data handling, certificate-based authentication, and an end-to-end secure data pipeline. A modular software architecture was developed to manage the full device lifecycle, including network registration, device provisioning, session coordination, secure data upload, and power-aware behavior. Complementing the SPU, the backend server provides certificate management (via STEP~CA), authenticated REST APIs, a session-control service, and encrypted data ingestion through an on-premises MinIO object-storage cluster. Combined, these components form a scalable, secure, and institution-compliant sensing infrastructure.

A series of evaluation experiments demonstrated that the proposed BSN meets its design objectives. The network-throughput experiment showed that the backend can reliably handle concurrent uploads from multiple SPUs without packet loss. Clock synchronization measurements confirmed that SPUs maintain sub-second alignment with the server, sufficient for temporal consistency in engagement analytics. Battery-runtime tests validated that the SPU can sustain over two hours of continuous operation, satisfying the demands of a typical lecture. Finally, inference benchmarks verified that the OAK-D Lite accelerator supports real-time execution of key CV/ML workloads, including emotion recognition, gaze estimation, and fatigue detection.

Overall, the contributions of this work establish a robust foundation for scalable, ethical, and real-time engagement sensing in university classrooms. The resulting BSN provides not only a practical tool for empirical research in student engagement but also a flexible platform for future extensions in learning analytics, classroom management, and adaptive instructional technology.

\section*{Future Work}

Although the system developed in this thesis is functional and demonstrates strong empirical performance, several avenues remain for future research and development:

\begin{itemize}
    \item \textbf{Improved Engagement Models:} Future versions may incorporate transformer-based vision models or multimodal fusion techniques to improve robustness and generalization.
    \item \textbf{Cross-Device Synchronization:} Hardware-triggered synchronization or precision time protocol (PTP) could further reduce temporal drift between SPUs.
    \item \textbf{Scalability Studies:} Large-scale deployments involving hundreds of SPUs would provide insights into long-term performance, network load, and maintenance requirements.
    \item \textbf{Adaptive Battery Management:} Predictive power modeling and dynamic frame-rate adjustment could extend battery life during long sessions.
    \item \textbf{Privacy-Preserving Analytics:} Techniques such as federated learning or on-device distillation could enable richer engagement inference while maintaining strict privacy guarantees.
\end{itemize}

In summary, this work demonstrates that a secure, autonomous, and ethically-compliant BSN is both technically feasible and practically deployable in real classrooms. The platform developed here opens the door to new research in real-time engagement analytics and has the potential to meaningfully transform how student learning is understood and supported at scale.



\backmatter

\bibliographystyle{osajnl}
\bibliography{dissertation}






\chapter{Curriculum Vitae}

\begin{center}
\large
Ahmed Elsayed
\end{center}

\bigskip


\section*{Education} 

\begin{itemize}
    \item \textbf{B.Sc. in Electronics and Communications Engineering}, 
    Zagazig University, Zagazig, Egypt \hfill Sept.~2011 -- Jun.~2016\\
    GPA: Excellent (87.73\%) \\
    Capstone Project: \emph{Talking Gloves} — an American Sign Language recognition system enabling communication for deaf individuals.
\end{itemize}

\section*{Professional Positions}

\begin{itemize}
    \item \textbf{Computer Vision Engineer}, Apple, Vision Pro Hands Team, 
    Sunnyvale, CA, USA \hfill Jan.~2023 -- Feb.~2024\\
    Led refactoring of critical dataset, training, metrics, and evaluation pipelines; 
    redesigned dataset-generation workflows (reducing creation time from 2 days to 30 minutes); 
    developed model-selection tools and label-correction frameworks; 
    optimized user-study labeling processes; 
    contributed extensively to the Sahara repository; 
    resolved FFMPEG frame mismatch bug, improving touch/no-touch label quality.

    \item \textbf{Graduate Research Assistant}, 
    Computer Vision and Image Processing (CVIP) Lab, University of Louisville \hfill Jan.~2022 -- Dec.~2023\\
    Projects include: 3D jaw reconstruction using an inter-oral sensor network; 
    real-time student engagement measurement in STEM classrooms (developed a MEAN-stack web application); 
    ATRV autonomous robot upgrades; 
    robotic arm system for autonomous vehicle refueling.

    \item \textbf{Hardware Engineer (Part-Time)}, 
    Egyptian Space Agency (EgSA) \hfill Oct.~2020 -- Dec.~2021\\
    Designed, simulated, and laid out the ZU-CubeSat communication subsystem PCB (UHF band).

    \item \textbf{Teaching Assistant}, 
    Faculty of Engineering, Zagazig University \hfill Nov.~2016 -- Dec.~2021\\
    Assisted in teaching: Electronic Devices, Analog Circuits, Integrated Circuits, RF Electronics. \\
    Led labs in: Electronic Circuit Design, HDL (Verilog/VHDL), PCB, IoT, Arduino, MATLAB.

    \item \textbf{Hardware \& Embedded Software Engineer}, Upwork Freelancing Platform \hfill May~2018 -- Apr.~2021\\
    Delivered PCB designs, embedded systems, IoT prototypes, high-speed digital boards, and DC-DC converters for global clients.

    \item \textbf{Cofounder \& Hardware Engineer}, ZagSystems Startup Company \hfill May~2018 -- Dec.~2021\\
    Designed hardware for IoT-based home automation and data-collection systems.
\end{itemize}

\section*{Publications and Presentations}

\begin{enumerate}
    \item \textbf{Samir Harb}; A. Elsayed; M. Yousuf; I. Alkabbany; A. Ali; S. Elshazley, 
    ``Accurate Colon Segmentation Using 2D Convolutional Neural Networks With 3D Contextual Information,'' 
    \textit{2024 IEEE International Conference on Image Processing (ICIP)}, 
    Abu Dhabi, United Arab Emirates, 2024, pp.~3212--3218, 
    doi: 10.1109/ICIP51287.2024.10647313.

    \item \textbf{Samir Harb}, M. Yousuf, A. Elsayed, A. Ali, S. Elshazly, and A. Farag,\\
    ``Deep Learning-Based Colon Segmentation for Accurate Colorectal Polyps Detection,'' 
    in \textit{Cancer Detection and Diagnosis}, CRC Press, 1st Edition, 2025, pp.~9.
\end{enumerate}

\end{document}